\documentclass[aps,pre,reprint,twocoloumn]{revtex4-1}

\usepackage{bm}
\usepackage{amsmath}
\usepackage[pdftex]{graphicx}
\usepackage[pdftex]{hyperref}

\begin{document}

\title{
  Stabilization of straight longitudinal dune under 
  bimodal wind with large directional variation
} 

\author{Sachi Nakao-Kusune}
\affiliation{Department of Physics, Kyushu University, Fukuoka
819-0395, Japan }

\author{Takahiro Sakaue}
\affiliation{Department of Physics and Mathematics, 
Aoyama Gakuin University,
5-10-1 Fuchinobe, Chuo-ku, Sagamihara, Kanagawa 252-5258, JAPAN}

\author{Hiraku Nishimori}
\affiliation{Department of Mathematical and Life Sciences, 
Hiroshima University, 
Higashi-Hiroshima, Japan}

\author{Hiizu Nakanishi}
\affiliation{Department of Physics, Kyushu University, Fukuoka
819-0395, Japan}

\date{\today}

\begin{abstract}
It has been observed that
  the direction in which a sand dune extends its crest line depends
  on seasonal variation of wind direction;
when the variation is small, 
  the crest line develops more or less
  perpendicularly to the mean wind direction 
  to form a transverse dune with some undulation.
  In the case of bimodal wind with a large relative angle, however,
  the dune extends its crest along the mean wind direction
  and evolves into an almost straight longitudinal dune.
  Motivated by these observations, we investigate the dynamical
  stability of isolated dunes using the crest line model, 
  where the dune dynamics is represented by its crest line motion.
First, we extend the previous linear stability analysis under the
  unidirectional wind to the case with non-zero slant angle
 between the wind direction and the normal direction of the crest line,
and show that the stability diagram  does not depend on the slant angle. 
Secondly, we examine how the linear stability is affected by the
  seasonal changes of wind direction in the case of bimodal wind with
  equal strength and duration.  For the transverse dune, we find that
  the stability is virtually the same with that for the unidirectional
  wind as long as the dune evolution during a season is small.
On the other hand, in the case of the longitudinal dune, the
  dispersions of the growth rates for the perturbation {are
  drastically different} from those of the unidirectional wind, and we
  find that the largest growth rate is always located at $k=0$. 
  This is because the growth of the perturbation with $k\ne 0$ is
  canceled by the alternating wind from opposite sides of the crest
  line even though it grows during each duration period of the bimodal
  wind.
For a realistic parameter set, the system is in the wavy unstable
regime of the stability diagram for the unidirectional wind, thus
the straight transverse dune is unstable to develop undulation and
  eventually evolves into a string of
barchans when the seasonal variation of wind direction is small, but
  the straight longitudinal dune is stabilized  under the large
  variation of bimodal wind direction.
%
We also perform numerical simulations on the crest line model, and
  find that the results are consistent with {our linear analysis
  and} the previous reports that show that the longitudinal dunes tend
  to have straight ridge and elongate over time.

\end{abstract}


\maketitle

\section{Introduction}

\bigskip
Aeolian sand dunes are natural patterns ubiquitous on
Earth\cite{Bagnold, Tsoar, Cooke, Lancaster}
and even on other planets {and satellites
\cite{Warren, Telfer-2018}}.
{Their time evolution involves several processes of sand
transport;  successive ballistic motion of sand due
to wind, called \textit{saltaion}, provides the main contribution to
the {windward (stoss) side} transport, while {\it avalanche}
is dominant on the {downwindward (leeward) } side because
the saltation is suppressed by the eddies formed at the crest line,
where the wind flow separates from the sand bed.
Lateral transport perpendicular to the wind
arises from \textit{reptation} \textit{i.e.} stochastic motion of the
grains ejected by collisions of saltation grains on the sand bed. }

{A flat sand bed under unidirectional wind destabilizes
in favor of sand wave due to the primary instability, which is now
well understood as a result of the phase lag between the flow shear
stress and the bed-form.  The minimum size of a dune is determined by
the  wavelength of the most unstable mode, which has been found to 
correspond with {the saturation length},
i.e., the length needed for the sediment to equilibrate with the
flow\cite{Andreotti-2002a, Andreotti-2002b, Kroy-2002a,
Kroy-2002b,Sauermann-2001}. 
{Under unsteady wind with varying strength
and direction, the secondary instabilities occur to develop
a variety of three-dimensional patterns, such as crescentic
barchans, star and linear dunes with complex and compound patterns
with a variety of length scales and orientations\cite{Pont-2014}.
Such instabilities are still mostly unexplored systematically.
}

From field observations, it has been noticed
that under seasonal
bimodal wind with comparable strength and small diversity angle, sand
bed develops into transverse dunes  with some undulation; its crest
lines extend perpendicular to the mean wind direction.
On the other hand, when the diversity angle of the bimodal winds is
sufficiently large, i.e. larger than 90 degrees, sand bed forms
longitudinal dunes, whose crest lines extend along the mean wind
direction\cite{Bagnold, Lancaster}.  
{
Such transition from the transverse to the longitudinal dune
has been also confirmed by
experiments\cite{Rubin-1987,Pont-2014,Rubin-1990,
Taniguchi-2012,Reffet-2010}
and simulations\cite{Parteli-2009,Parteli-2007}.
Understanding the selection mechanism of
}
emerging dune types by the pattern of the wind direction change 
has been recognized as one of fundamental issues of dune
morphology.\cite{Bagnold, Tsoar, Cooke, Lancaster, Warren}.

In addition to the wind direction, 
{
if the wind strength and the duration also vary in the bimodal
wind,}
the relation between the wind pattern and the
crest line direction of dune should become far more complicated. 
It is notable that such a relationship may be summarized
by a simple empirical hypothesis called ``the maximization
of the gross sand transport normal to the crest'' \cite{Rubin-1987,
Rubin-1990}.  The theoretical understanding 
of this hypothesis, however, has not been reached yet,
and except for a few approaches\cite{Werner-1997,Gadal-2019}, little has been
worked out theoretically to clarify how the dune direction is
determined under the unsteady wind.
The difficulty is that most of theoretical descriptions of dune
evolution are
based on complex formulation to describe mixed effects of fluid
and granular dynamics, thus analytical treatment is not 
feasible even for boldly simplified numerical models like cell
models\cite{Werner-1995, Nishimori-1993, Nishimori-1998}.  

An exception is the crest line model, which describes the dune
evolution by a couple of partial-differential equations for the
position and height of the crest line of the dune\cite{Guignier-2013}.
{Its dynamics is based on simple assumptions for the longitudinal
and the lateral transportations of sands.}
Despite its simplicity, it has been demonstrated that the
model is capable to reproduce basic feature of the dune dynamics with
its instability\cite{Guignier-2013}.
In this paper, using the crest line model, we present theoretical
analysis and numerical simulations for the stability of
dune dynamics under bimodal winds with equal strength and duration in
the case of the non-zero slant angle $\chi$, i.e. in the case where the
wind direction is not normal to the crest line. 

The paper is organized as follows.  After introducing the crest
line model in Sec.2, the linear stability of the straight crest with
the non-zero slant angle $\chi$ is presented in Sec.3 to show that the
stability diagram does not depend on $\chi$.
Then in Sec.4, the formalism of the linear stability analysis is
extended to the case for the bimodal wind with the equal wind strength
and the equal duration.  The expressions for the growth rate of
perturbation are obtained for the case of the transverse dune, 
and for the case of the longitudinal dune. 
In Sec.5, these growth rates are numerically estimated as a function
of the wave number $k$ for various slant angle $\chi$ and parameter
sets, and
in Sec.6, numerical simulations are presented for the crest line model
to compare with the results of the linear stability analysis.
The concluding remarks are given in Sec.7.

\section{Crest line model}

The crest line model is originally introduced by Guignier \textit{et
al.}\cite{Guignier-2013} to describe dune dynamics under
unidirectional steady wind. A dune on a non-erodible bed is
represented by its crest line with its stream-wise position $x$ and
height $h$ as a function of $y$, the position along the co-ordinate
axis perpendicular to the streamline (Fig.\ref{fig:co-ordinate}(a)).
Following the study of Niiya et al.\cite{Niiya-2010,Niiya-2012}, the
key hypothesis for simplification in this model is that any
stream-wise cross-section of dune should be similar triangle.
Then, the dune configuration at the time $t$ is uniquely defined by the
position $x(y,t)$ and the height $h(y,t)$ of the crest
with the angles of the windward and
downwindward slopes, $\theta$ and $\phi$.

\begin{figure}[b]
  \centerline{
\includegraphics[width=80mm]
  {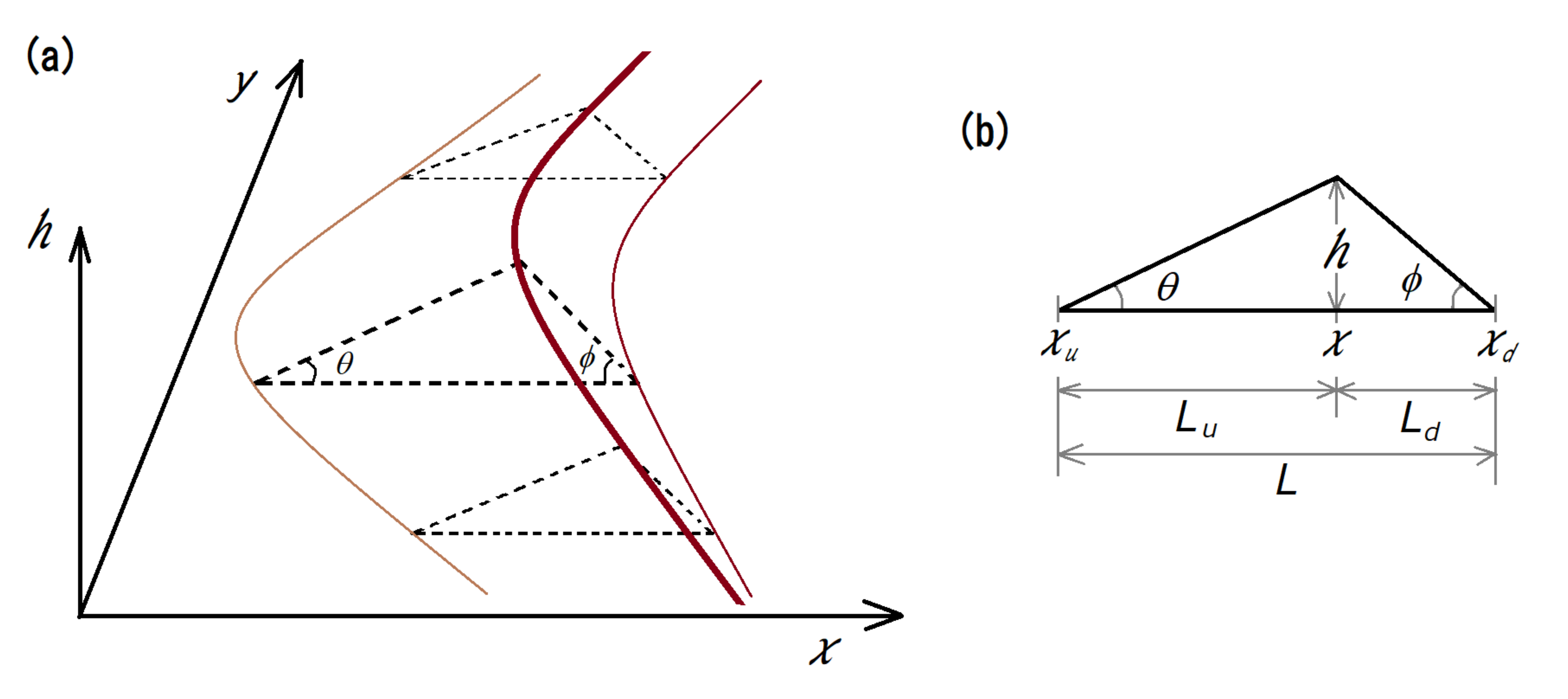}
}

\caption{Co-ordinate system for the crest line model and similar
  triangle of dune cross section. 
  (a) The $x$-axis is taken to be parallel to the wind direction and
  the $y$-axis is perpendicular to it; the $h$-axis is vertical.  The
  crest line of the sand dune is shown by the thick brown line, which
  is represented by $(x(y), y, h(y))$. 
  cross section of the dune by a plane parallel to the $x$-$h$ plane
  is similar triangle, whose peak is at the crest line.  The remaining
  two vertices on the base of the triangle are the windward and
  downwindward ends of the dune, which are shown by the thin brown
  lines, $(x_u(y),y)$ and $(x_d(y),y)$ in the $x$-$y$ plane.  (b) The
  height and the length of the base of the triangle are $h$ and $L$.
  $L_u$ and $L_d$ are the lengths of the windward and the
  downwindward parts of the base.  }
  \label{fig:co-ordinate}
\end{figure}

It is convenient to introduce 
the positions of the windward and the downwindward ends of the slope,
$x_u$ and $x_d$, respectively, which are given by
\begin{equation}
  x_u =  x - {B\over A} h, \qquad
  x_d =  x + {C\over A} h.
  \label{eq:x_u-x_d}
\end{equation}
The parameters $A$, $B$, and $C$ are
the geometrical {parameters} defined as
\begin{eqnarray}
A\equiv\frac{h}{L},\quad
B\equiv\frac{L_u}{L}, \quad
C\equiv\frac{L_d}{L}, 
  \label{eq:ABC}
\end{eqnarray}
where
$L$, $L_u$, and $L_d$ are the total, the windward side, and the
downwindward side lengths of the dune, respectively
(Fig.\ref{fig:co-ordinate}(b)).
Note that the simple equality
\begin{equation}
B + C = 1
\end{equation}
holds because $L=L_u + L_d$.

\begin{figure}
\begin{center}
\includegraphics[width=60mm]{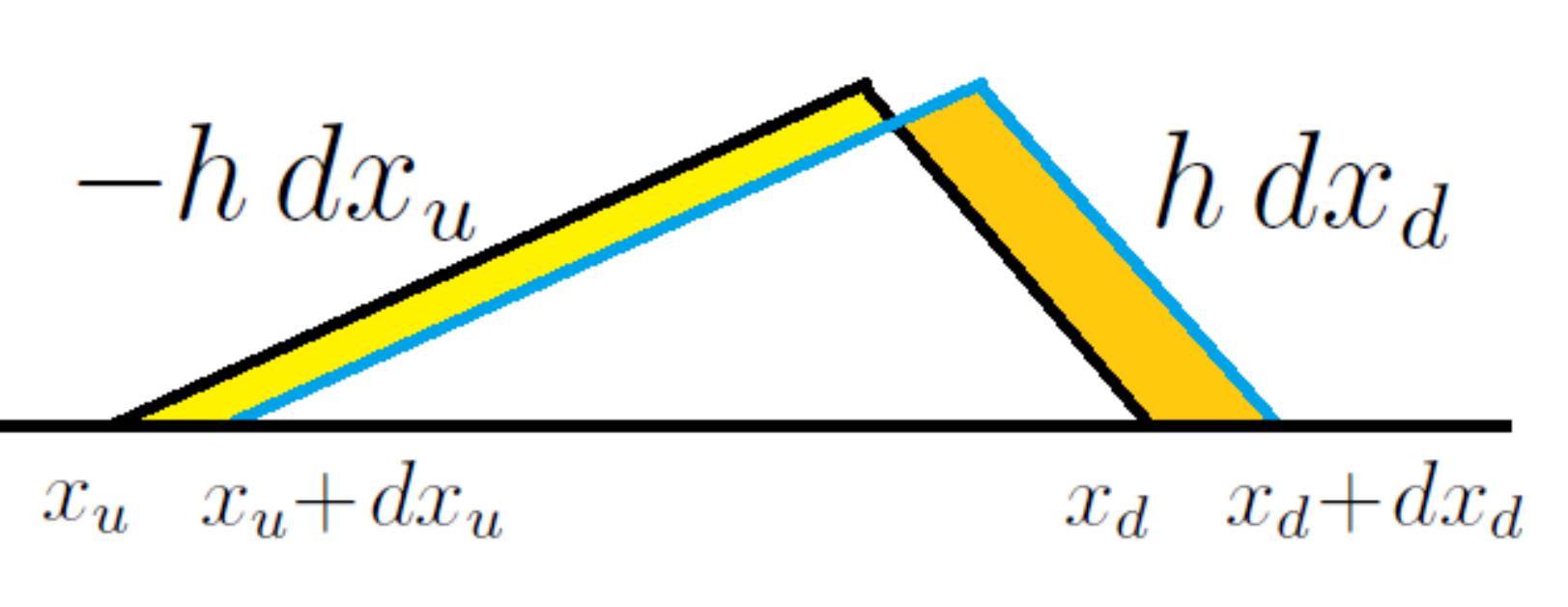}
  \end{center}
  \caption{
    Change of the cross section caused by small advance of the sand
  dune. The total change of the cross section consists of the
  decrement $h\,dx_u$ in the windward side and the increment
  $h\,dx_d$ in the downwindward side.
  }
\label{fig:cross-section}
\end{figure}

The time development of the crest line is determined by the sand
transport in the $x$ and the $y$ directions, i.e. the
longitudinal transport and the lateral transport.
The longitudinal transport {caused by wind} consists of the
incoming sand flux {density} $f_{\rm in}$ and the outgoing flux
{density} over the crest line  $q$.  Both depend on the flow
strength, and are determined by the upwind boundary.
A fraction of the sand flux $q$ over the crest line is captured on the
downwind slope; that fraction is called as the efficiency rate
$T_e$\cite{Cooke, Momiji-2000} and
shown to be an increasing function of the dune height $h$ for a
given wind strength\cite{Momiji-2000}.

{The lateral transport is caused by random motion of sand grains
biased by lateral gradient of the dune face.}
%
%
{Thus},  the local lateral flux densities are assumed to be 
proportional to the gradients of the dune
faces in the transverse direction  with the diffusion coefficients
$D_u$ and $D_d$ of the upwind and the downwind slope, respectively.
Then the total lateral fluxes 
{on the windward and the downwindward slopes}, 
$J_u$ and $J_d$, integrated over each
dune face along the longitudinal direction are shown to be
\begin{eqnarray}
  J_u&=&  D_u\, h{\partial x_u\over\partial y},
\label{eq:Ju}\\
J_d&=&  -D_d\, h{\partial x_d\over\partial y}.
\label{eq:Jd}
\end{eqnarray}

The change of the cross section by the displacement of the upstream
slope is $-hdx_u$ and that by that of the downstream slope is $h
dx_d$ (Fig.\ref{fig:cross-section}).  By balancing these changes with
the influxes of grain through the corresponding slopes of the dune, we
obtain the time evolution equations
\begin{align}
  -h {\partial x_u\over\partial t} 
  & = f_{\rm in} - q - {\partial J_u\over\partial y},
  \label{eq:x_u}
  \\
  h {\partial x_d\over\partial t} 
  & = T_e q - {\partial J_d\over\partial y}.
  \label{eq:x_d}
\end{align}
Using Eqs.(\ref{eq:x_u-x_d}), these equations lead to
the time evolution for $x$ and $h$,
\begin{eqnarray}
{1\over A}\,h\frac{\partial h}{\partial t}
&=&
f_{\rm in}-(1-T_e)q - \frac{\partial J_d}{\partial y} -
\frac{\partial J_u}{\partial y},
\label{eq:h}
\\
h\frac{\partial x}{\partial t}
&=&
 q(BT_e+C)-Cf_{\rm in} - B\frac{\partial J_d}{\partial y}
          +C\frac{\partial J_u}{\partial y},
\label{eq:x}
\end{eqnarray}
which are Eqs. (3) and (4) in ref.\cite{Guignier-2013}.

\begin{figure}
\centerline{\includegraphics[width=4.0cm]{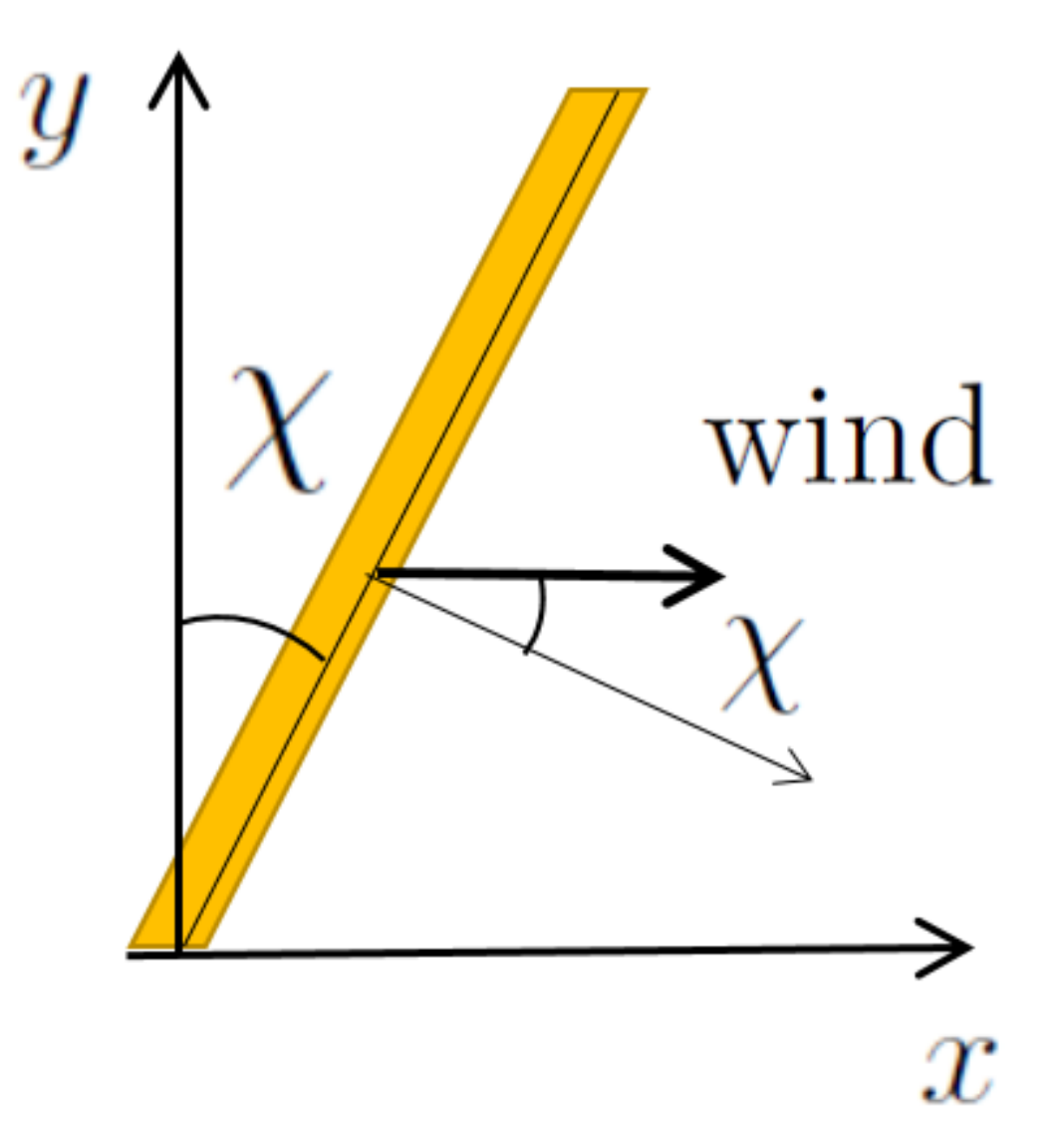}}
  \caption{Straight dune with the slant angle $\chi$.
  The wind direction is shown by the thick arrow, which is slanted by
  the angle $\chi$ from the normal direction of the crest line.}
  \label{fig:unidirectional-wind}
\end{figure}
Eqs. (\ref{eq:h}) and (\ref{eq:x}) have the steady uniform solution,
\textit{i.e.} the infinite rectilinear dune of the height $h_0$
migrating with the constant {speed} $V_0$ with the slant angle
$\chi$,
\begin{eqnarray}
x_0(y,t)&=&V_0 t+y\tan\chi,
\label{eq:teijo-x}
\\ 
h_0(y,t)&=&h_0,
\label{eq:teijo-h}
\end{eqnarray}
as shown in Fig.\ref{fig:unidirectional-wind}.
Here, $h_0$ and $V_0$ are determined by
\begin{eqnarray}
0 & = & f_{\rm in} - \Bigl( 1 - T_e (h_0)  \Bigr)\, q,
\label{eq:teijo-h_0}\\
V_0&=&\frac{qT_e(h_0)}{h_0} = \frac{q-f_{\rm in}}{h_0}.
\label{eq:teijo-V_0}
\end{eqnarray}
The migration {speed} $V_0$  is inversely proportional to the height
of the dune, but both $h_0$ and $V_0$ are independent of the slant
angle $\chi$.
This comes from the fact that $q$ and $T_e$ are {taken to be}
independent of $\chi$ or $\partial x/\partial y$, {which should be
reasonable approximation for small $\chi$ because of the
$\chi\leftrightarrow -\chi$ symmetry of the problem, but may not be
valid for large value of $\chi$.
Note that the migration {speed} $V_0$ by Eq.(\ref{eq:teijo-V_0})
is the {speed} along the wind direction, thus the speed normal the
crest line is given by $V_0\cos\chi$, which is smaller than $V_0$.}

For numerical simulations in Sec.\ref{Sec:numerics}, 
we take $f_{\rm in}=0$ and assume that $T_e$ and $q$ depend on $h$ as
\begin{align}
  T_e(h) & = \left\{\begin{array}{ll}
    h/h_c \qquad & \mbox{for $0\le h\le h_c$} \\
    1 & \mbox{for $h>h_c$}
  \end{array}\right.
  \label{T_e-h}
  \\
q(h) & = \left\{\begin{array}{ll}
   q_c\; h/h_c \quad& \mbox{for $0\le h\le h_c$} \\
    q_c & \mbox{for $h>h_c$}
  \end{array}\right.  
  \label{q-h}
\end{align}
with a critical height $h_c$.
{
The form $T_e(h)$ of Eq.(\ref{T_e-h}) is employed by Ref.
\cite{Guignier-2013} as the simplest form for a monotonically
increasing function with the saturation value 1, which 
means that dunes can be self-sustained
without the influx $f_{\rm in}$ if it is large enough.  We
introduce the form $q(h)$ of Eq.(\ref{q-h}) here because $q$ should be
zero in the $h\to 0$ limit when $f_{\rm in}=0$.  
The critical heights $h_c$ for these quantities are not
necessarily the same, but would be of the same order
related to the saturation length of the sand transport.
We take them the same for simplicity.
}

%
\section{Linear Stability of rectilinear dunes for non-zero $\chi$}

In this section, we present the linear stability analysis of the steady
uniform solution Eqs.(\ref{eq:teijo-x}) and (\ref{eq:teijo-h}) for
non-zero value of $\chi$.
As for the case with $\chi=0$, 
i.e. the case of the transverse dune with
the crest line  perpendicular to the wind,  
the stability analysis has been worked out
by Guignier \textit{et al.\ }\cite{Guignier-2013}, 
and the stability has been determined in terms of two
dimensionless
parameters,
\begin{equation}
\delta = \frac{h_0T_e'(h_0)}{T_e(h_0)},\qquad
\rho = \frac{D_d}{D_u}.
\end{equation}
The positive $\delta$ destabilizes the steady solution because 
the higher dune height $h$ increases the sand deposit on the downwind
slope $qT_e(h)$, which makes the dune height even higher.
The larger $\rho$ also destabilizes it because the lateral diffusion
in the downwindward slope accumulates the sand in the concave crest
line region, which further delay the advance of the crest line, while
the lateral diffusion in the windward slope has the opposite effect.
They obtained the stability diagram in the $\delta-\rho$ plane.

We extend the linear stability analysis to the case
with the non-zero $\chi$ and the case under the bimodal winds with
the same strength and duration.

\subsection{Linear analysis}

The stability of the steady solution Eqs.(\ref{eq:teijo-x}) and
(\ref{eq:teijo-h}) are examined by the sinusoidal perturbation
of the wave number $k$ with the
coefficient $x_k(t)$ and $h_k(t)$,
\begin{eqnarray}
  x(y,t) & = & x_0(y,t) + x_k(t) e^{iky},
\label{eq:xx}
\\
h(y,t) & = &h_0+h_k(t) e^{iky}.
\label{eq:hh}
\end{eqnarray}
Note that $k$ is the wave number not along the crest line but
along the $y$-axis.
Substituting Eqs. (\ref{eq:xx})
and (\ref{eq:hh}) into Eqs. (\ref{eq:h}) and (\ref{eq:x}),
within the linear approximation we obtain
\begin{equation}
  {d\over dt}  \left|\bm x_k(t)\right>
    =
    \hat A_k(\chi)   \left|\bm x_k(t)\right>
    \label{eq-for-x_k-h_k}
\end{equation}
with the ket vector
\begin{equation}
  \left|\bm x_k(t)\right> \equiv
  \left(\begin{array}{c} x_k(t) \\ h_k(t)\end{array}\right),
\end{equation}
and the matrix $\hat A_k(\chi)$ given by
\begin{widetext}
\renewcommand{\arraystretch}{1.5}\begin{align}
   \hat A_k(\chi) \equiv
  & \left(\begin{array}{cc}
    -(BD_d+CD_u)k^2, &\displaystyle
 {qBT_e'-V_0\over h_0}+{BD_d+CD_u \over h_0}\, ik\tan\chi
    -{BC\over A}(D_d-D_u)k^2 \\
    -A(D_d-D_u) k^2, &\displaystyle
   {AqT_e'\over h_0}+{A(D_d-D_u)\over h_0}\,ik\tan\chi - (CD_d+BD_u)k^2
  \end{array}\right).
  \label{A_k(chi)}
\end{align}
\renewcommand{\arraystretch}{1}
\end{widetext}

For later use, we introduce the notation for the eigenvalues
$\omega_k^\pm(\chi)$ 
(${\rm Re}\,\omega_k^+(\chi)>{\rm Re}\,\omega_k^-(\chi)$) and the
corresponding left and right eigenvectors $\left<\bm
x_k^\pm(\chi)\right|$ and $\left|\bm x_k^\pm(\chi)\right>$,
respectively, with the ortho-normalization conditions
\begin{equation}
  \left<\bm x_k^\pm(\chi)|\bm x_k^\pm(\chi)\right>=1, \quad
  \left<\bm x_k^\pm(\chi)|\bm x_k^\mp(\chi)\right>=0.
\end{equation}
The eigenvalues $\omega_k^\pm(\chi)$ are determined by
the characteristic equation,
\begin{widetext}
\begin{equation}
  \omega^2 - \omega\left[
    F(k) + {A\over h_0}(D_d-D_u)ik\tan\chi \right] + G(k) k^2 = 0,
  \label{char-eq-chi}
\end{equation}
as
\begin{align}
  \omega_k^\pm(\chi) & =
  {1\over 2}\left(
  F(k)+{A\over h_0}\big(D_d-D_u\big) ik\tan\chi 
  \pm
  \sqrt{\left(
  F(k)+{A\over h_0}\big(D_d-D_u\big) ik\tan\chi \right)^2
  -4 G(k) k^2 }\; \right)
  \label{omega_k^pm(chi)}
\end{align}
\end{widetext}
with the functions $F(k)$ and $G(k)$ defined by
\begin{align}
  F(k) & \equiv {AqT_e'\over h_0} - (D_d+D_u)  k^2,
  \label{F(k)}
  \\
  G(k) &
  \equiv -{A\over h_0}\big( D_uqT_e' + V_0(D_d-D_u)\big) + D_d D_u k^2.
  \label{G(k)}
\end{align}
Note that Eqs.(\ref{char-eq-chi})$\sim$(\ref{G(k)}) do not depend on
$B$ and $C$ even though Eq.(\ref{A_k(chi)}) depends on them.

\subsection{Stability diagram for unidirectional wind with non-zero $\chi$}

In the case of unidirectional wind, we demonstrate that the stability
diagram for $\chi\ne 0$ is the same as that for $\chi=0$ as is shown
in Fig.~\ref{fig:nanamesouzu}.  For the the uniform perturbation with
$k=0$, it is easy to see that the stability is the same with the case
of $\chi=0$ because the $\chi$ term always appears in combination with
the wave number $ik$.  In this case, the characteristic equation
(\ref{char-eq-chi}) has the solutions
\begin{equation}
  \omega = {AqT_e'\over h_0},\quad 0,
  \label{omega-for-k=0}
\end{equation}
therefore the uniform solution given by (\ref{eq:teijo-x}) and
(\ref{eq:teijo-h}) is 
unstable for $T_e'>0$ (or $\delta>0$) and
marginally stable for $T_e'\le 0$ (or $\delta\le 0$).

The stability against the $k\ne 0$ perturbation can be determined by
examining the dispersion of the unstable mode of
Eq.(\ref{omega-for-k=0}) for $T_e'>0$,
or the marginally stable mode of Eq.(\ref{omega-for-k=0}) for $T_e'<0$.
In the region $\delta>0$ ($T_e'>0$), the dispersion of the unstable mode is
expanded as
\begin{widetext}
\begin{equation}
  \omega 
  = {AqT_e'\over h_0} + {A\over h_0}(D_d-D_u) ik\tan\chi
  + \left({V_0\over qT_e'}(D_d-D_u) -D_d\right) k^2 + O(k^3),
\end{equation}
\end{widetext}
for small $k$,
thus the $k\ne 0$ mode is even more unstable than that of $k=0$ if
\begin{equation}
  D_d > {D_u\over 1-qT_e'/V_0}\quad\mbox{or}\quad
  \rho > {1\over 1-\delta}.
  \label{boundary-1}
\end{equation}
The dispersion for the marginal mode in the $\delta<0$ ($T_e'<0$) region
is expanded as
\begin{equation}
  \omega = - \left( D_u + {V_0\over qT_e'} (D_d-D_u)\right) k^2
  + O(k^3),
\end{equation}
for small $k$,
thus the $k\ne 0$ mode is unstable for
\begin{equation}
  D_d > D_u \left(1-{qT_e'\over V_0}\right)\quad\mbox{or}\quad
  \rho > 1-\delta.
  \label{boundary-2}
\end{equation}
All of these stability conditions, Eqs.(\ref{omega-for-k=0}),
(\ref{boundary-1}), and (\ref{boundary-2}),  do not depend on the angle
$\chi$, therefore, {\it the stability diagram given by
Fig.\ref{fig:nanamesouzu} for the uniform solution is
the same with the case of $\chi=0$}.

\begin{figure}[bt]
  \begin{center}
    \includegraphics[width=70mm]{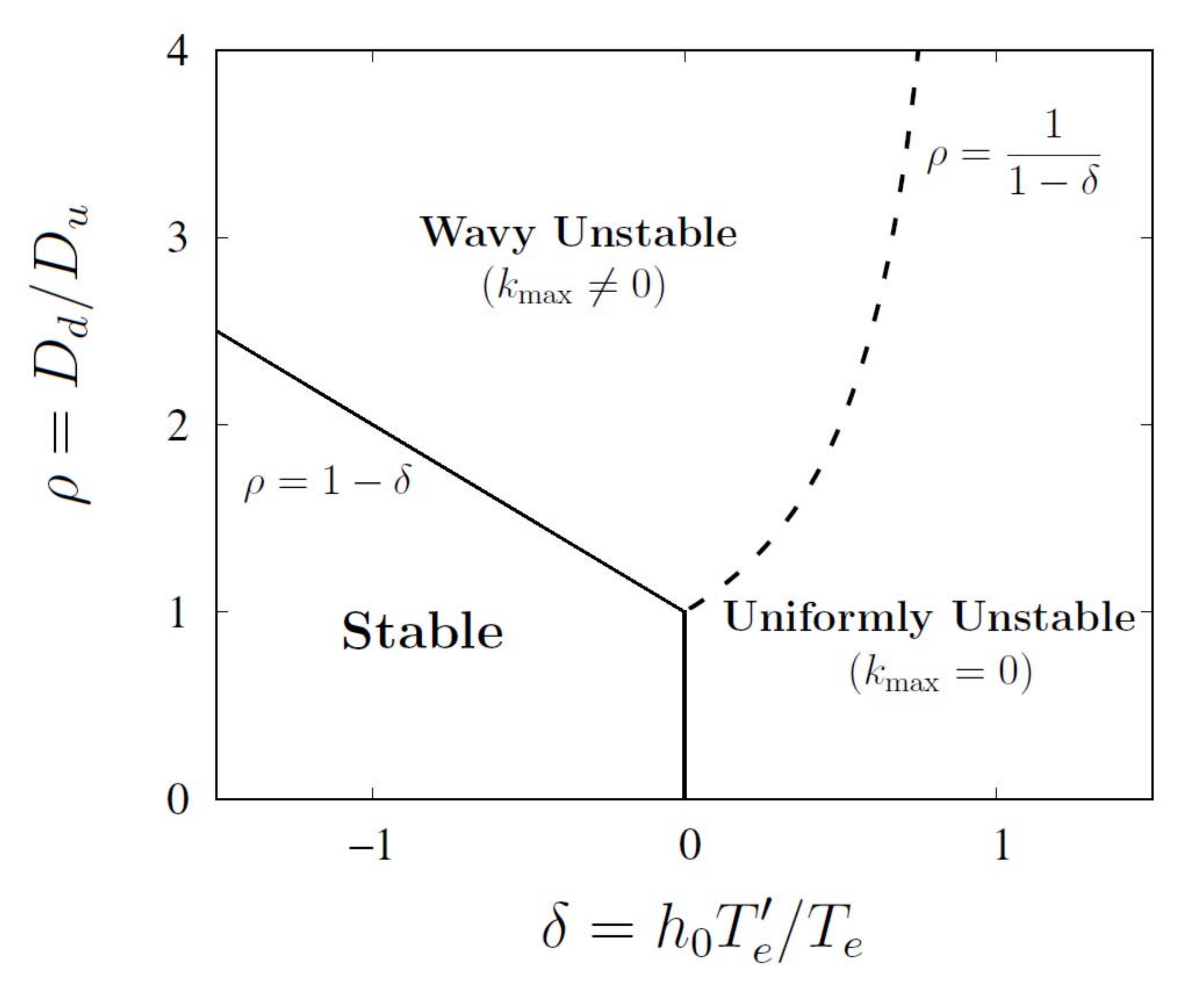}
  \end{center}
  \caption{Stability diagram for both $\chi=0$ and $\chi\ne 0$ cases.}
  \label{fig:nanamesouzu}
\end{figure}

There are three regimes of stability: (i) {\it stable regime},
\textit{i.e.} the steady solution is linearly stable for any $k$,
(ii) {\it uniformly unstable regime}, \textit{i.e.} there exists
a band of unstable modes with the most unstable mode at $k_{\rm max}=0$,
and (iii) {\it wavy unstable regime}, \textit{i.e.} there exists a band of
unstable modes with the most unstable mode at $k_{\rm max}\ne 0$.
Although the boundaries of these regimes do not depend on $\chi$,
the wave number $k_{\rm max}$ of the most unstable mode  depends on $\chi$
in the wavy unstable regime.

In the real aeolian sand dunes, avalanches in the downwindward slope
dominate the lateral sand transport, which suggests $\rho\gg 1$.  By
{a numerical estimation using field data, Momiji \textit{et
al.\ }\cite{Momiji-2000}} showed that the efficiency rate $T_e$ for aeolian dunes is an
increasing function of the dune height ($\delta>0$).  The parameters of
the real aeolian sand dunes, therefore, are presumed to be in the wavy
unstable regime.

\section{Linear stability of rectilinear dune under bimodal winds}
\begin{figure}[bt]
  \begin{center}
    \includegraphics[width=70mm]{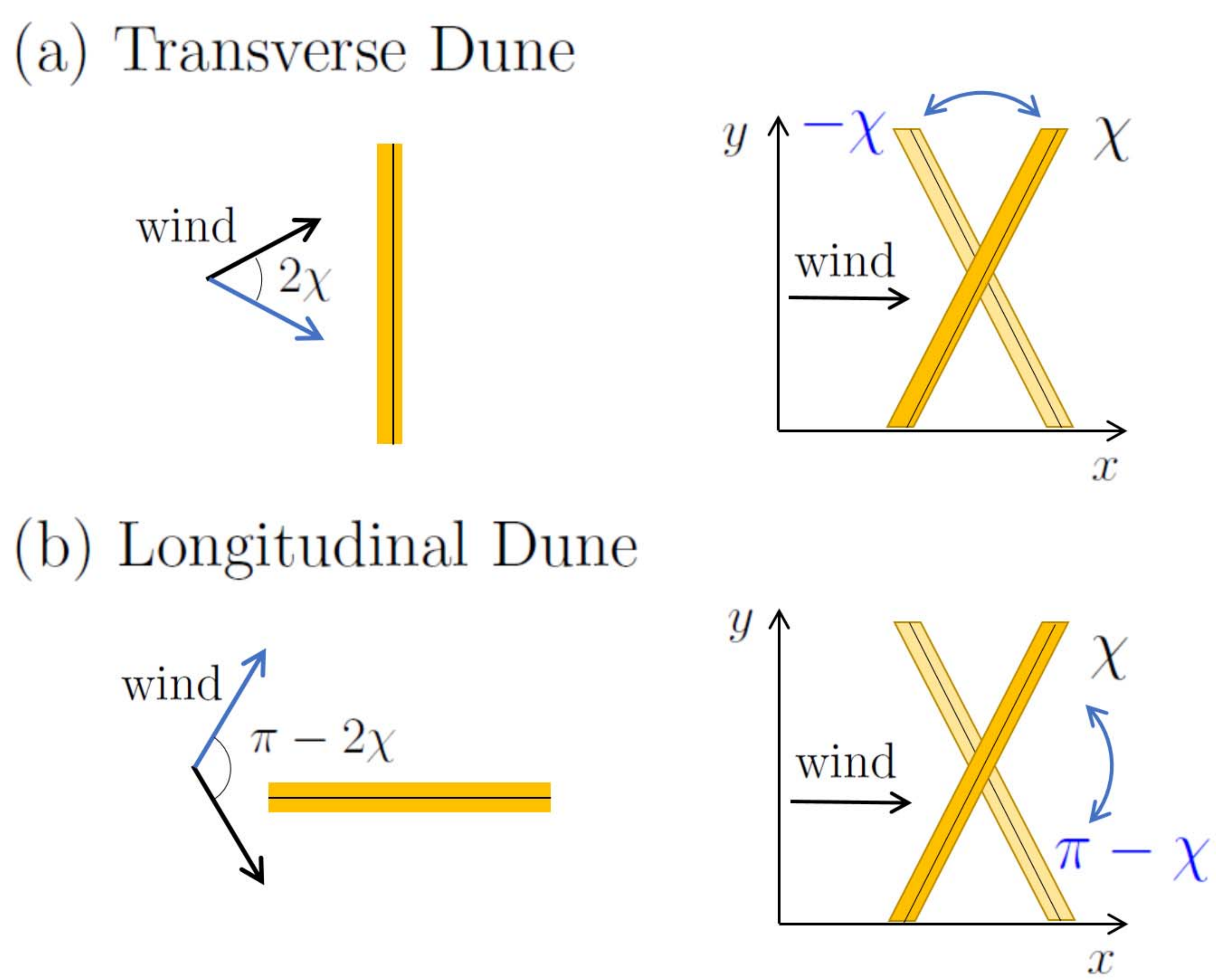}
  \end{center}
  \caption{
Transverse dune (a) and longitudinal dune (b) under bimodal wind.
The wind directions are shown by the black and blue arrows.
In the crest line model, the $x$-axis is taken parallel to the wind
  direction, therefore, 
  every time the wind changes its direction,
  the crest line need to be rotated between the two slant angles, i.e.
  $\chi$ and $-\chi$ for the transverse dune, and $\chi$ and
  $\pi-\chi$ for the longitudinal dune.
}
  \label{fig:bimodal-wind}
\end{figure}

Now, we extend the above stability analysis to the bimodal wind case,
where the wind direction changes alternately.
We consider the two cases: (A) the case of transverse dune, where the
wind direction switches between $\chi$ and $-\chi$, and (B) the
case of longitudinal dune, where the wind direction changes between
$\chi$ and $\pi-\chi$ (Fig.\ref{fig:bimodal-wind}).
The angles between the bimodal winds are $2\chi$ for the transverse
dune and $\pi-2\chi$ for the longitudinal dune.
The wind is supposed to blow in each direction for the same duration
time $T$.
We also assume that the geometrical {parameters} $A$, $B$, and $C$
relax quickly to the original steady values after the wind direction
switches even in the case where the windward and the downwindward
sides are interchanged.
{
This is not necessarily true in the usual situation, where the
migration distance during the duration time of each wind is shorter
than the width of the dune.  In such a situation, the geometrical
{parameters}, $A$, $B$, and $C$ should relax to some averaged values over
many periods.
}

\subsection{Transverse dune}

First, we consider the transverse dune under the bimodal wind with the
direction $\chi$ and $-\chi$.  In our formalism (\ref{eq:x}) and
(\ref{eq:h}), the wind is always supposed to be in the $x$-direction
and the uniform solution (\ref{eq:teijo-x}) and (\ref{eq:teijo-h}) is
inclined by the angle $\chi$, thus we rotate the solution by
$\pm 2\chi$ as in Fig.\ref{fig:bimodal-wind}(a) when the wind
direction switches, which corresponds that the co-ordinate system is
rotated by $\mp 2\chi$.

Suppose that the wind blows in the $\chi$ direction for $0<t<T$, then
the time evolution of the perturbation is governed by
Eq.(\ref{eq-for-x_k-h_k}) and the solution is given by
\begin{equation}
\left|\bm x_k(t)\right> 
  = e^{\hat A_k(\chi) t}  \left|\bm x_k(0)\right>.
\end{equation}
At $t=T$, the wind direction changes from $\chi$ to $-\chi$.  We
change the zero-th order solution (\ref{eq:teijo-x}) and
(\ref{eq:teijo-h}) by replacing $\chi$ by $-\chi$, but the
perturbation $x_k$ and $h_k$ remain the same.
Thus the perturbation evolve for $T<t<2T$ as
\begin{equation}
\left|\bm x_k(t)\right> 
= e^{\hat A_k(-\chi) (t-T)}  e^{\hat A_k(\chi) T}  \left|\bm x_k(0)\right>,
\end{equation}
therefore, after one cycle of the bimodal wind at $t=2T$, the perturbation
evolves as
\begin{equation}
  \left|\bm x_k(2T)\right> 
  =  \hat\Lambda_{\rm trans}  \left|\bm x_k(0)\right>
\end{equation}
with the evolution matrix
\begin{equation}
  \hat\Lambda_{\rm trans}
  \equiv e^{\hat A_k(-\chi) T}  e^{\hat A_k(\chi) T}
  = 
  \left( e^{\hat A_k(\chi) T}\right)^*\,  e^{\hat A_k(\chi) T}.
  \label{Lambda_trans}
\end{equation}
The stability of the transverse dune under the bimodal wind can be
determined by the eigenvalues $\lambda_{\rm trans}^\pm$ of the matrix
$\hat\Lambda_{\rm trans}$.
We define 
\begin{equation}
  \bar{\omega}_{\rm trans}^\pm 
  \equiv {\log \lambda_{\rm trans}^\pm \over 2T},
  \label{ave_omega_trans}
\end{equation}
then the average growth rate over the one period of the bimodal wind is
given by ${\rm Re\,}[\bar{\omega}_{\rm trans}^\pm ]$.

\subsection{Longitudinal dune}

In the case of longitudinal dune, the analysis can be performed almost
in the same way as in the case of the transverse dune, but the
difference is that the windward and the downwindward sides are
interchanged when the wind direction changes between $\chi$ and
$\pi-\chi$, thus $x_k$ should be replaced by $-x_k$ while $h_k$
remains the same. This can be represented by
\begin{equation}
  -\hat\sigma_z \left| \bm x_k \right>
\end{equation}
with the spin matrix
\begin{equation}
  \hat\sigma_z = 
  \left(\begin{array}{cc} 1,&0\\ 0, &-1\end{array}\right).
\end{equation}
At the same time, 
the sign of wave number of the perturbation is also inverted
because the direction of the $y$ axis flips as can be seen in
Fig.\ref{fig:bimodal-wind}(b).
Thus, the time evolution matrix for the one period of the bimodal wind
is given by
\begin{equation}
  \hat\Lambda_{\rm long}
  \equiv 
  (-\hat\sigma_z)e^{\hat A_{-k}(\pi-\chi)T}
  (-\hat\sigma_z) e^{\hat A_k(\chi)T}
  =
  \left(  \hat\sigma_z\,  e^{\hat A_{k}(\chi)T}\right)^2
  \label{Lambda_long}
\end{equation}
and the stability is determined by its eigenvalues $\lambda_{\rm
long}^\pm$ or the average growth rate, i.e. the real part of 
 $\bar{\omega}_{\rm long}^\pm$ defined as
\begin{equation}
  \bar{\omega}_{\rm long}^\pm 
  \equiv {\log \lambda_{\rm long}^\pm \over 2T}.
  \label{ave_omega_long}
\end{equation}

\section{Numerical estimate of growth rate}\label{numerical_growth}

\begin{figure}
  \centerline{
    \includegraphics[width=9cm]{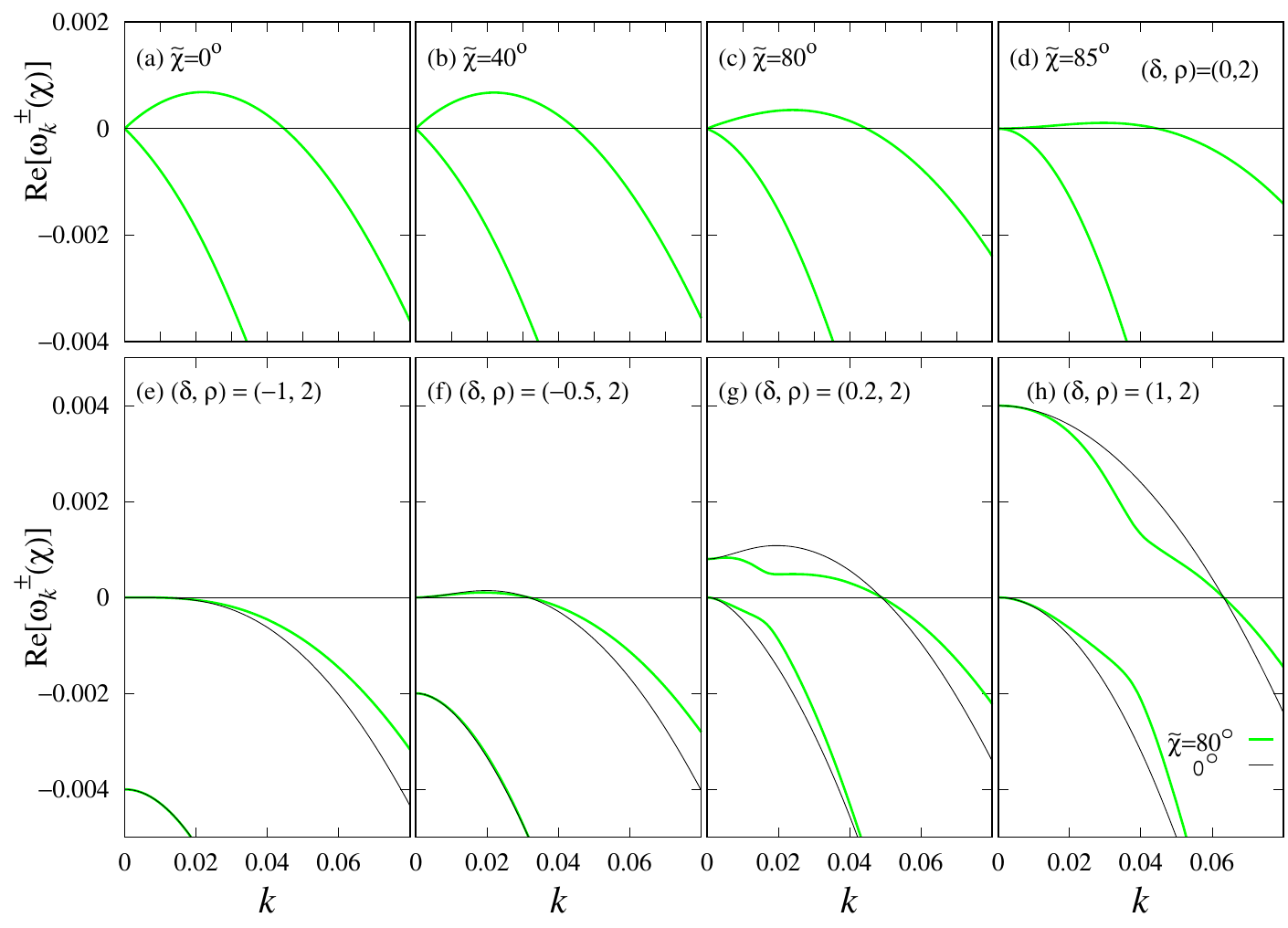}}
\caption{The dispersions of the growth rates 
${\rm Re}\,[\omega_k^\pm(\chi)]$ of the perturbation
{under the unidirectional wind}
for some values of $\delta$ and the wind direction {$\tilde\chi$}
  in the wavy unstable region with $\rho=2$.  The parameters are
  $h_0=5$, $q=1$, $D_u=1$, $T_e(h_0)=1$, $(A,B,C)=(0.1,0.8,0.2)$ with
  $T_e'(h_0)=(T_e(h_0)/h_0) \delta$ and $D_d=D_u\rho$.}
\label{fig:growth_rate-uni}
\end{figure}

We numerically estimate
the average growth rates of the sinusoidal perturbation to the uniform
solution with non-zero slant angle.
To represent numerical values,
we employ the unit system where
\begin{equation}
  h_c = q_c = D_u = 1;
  \label{unit_system}
\end{equation}
{$h_c$ and $q_c$ are 
the critical height and the saturated flux, respectively, 
in Eqs.(\ref{T_e-h}) and (\ref{q-h}), that define the functions $T_e(h)$
and $q(h)$.
We adopt the same unit for $h$ and $x$, then
the numerical values for $h$, $x$, $y$, and $t$ should read as those for
\begin{equation}
  {h\over h_c}, \quad {x\over h_c}, \quad 
  {y\over h_c \sqrt{D_u / q_c}}, \quad 
  {t\over {h_c^2/ q_c}},
  \label{unit}
\end{equation}
respectively\footnote{{
Since $h$, $x$, and $y$ have the same dimension of unit, there are
actually three possible ways to interpret the numerical values
depending on the choice of two quantities out of the three to be
assigned to the same unit.  We prefer $h$ and $x$ to be in the same
unit because the geometrical parameters $A$, $B$, and $C$ do not
depend on the constants of Eq.(\ref{unit_system}).  
In this choice, however, the slant angle
$\chi$ depends on $\sqrt{D_u/q_c}$ as in Eq.(\ref{tilde_chi}).
}}.
In this system, numerical values of the slant angle should be given
by $\tilde\chi$ defined by
\begin{equation}
  \tan\tilde\chi \equiv \sqrt{D_u\over q_c}\,\tan\chi
  \label{tilde_chi}
\end{equation}
from the actual slant angle $\chi$
because the unit is different between the $x$ and $y$
direction\footnote{{
  In the following, numerical values for the slant angle are always
given by $\tilde\chi$ defined in Eq.(\ref{tilde_chi}), but the slant
angle in the mathematical expressions should be the original slant
angle $\chi$.
}}.
}

Considering the case of {saturated self-sustained dune}
\begin{equation}
  f_{\rm in}=0, \quad T_e(h_0) = 1, \quad q=q_c,
\end{equation}
the perturbation is examined around the uniform solution
(\ref{eq:teijo-x}) and (\ref{eq:teijo-h}) with
\begin{equation}
  h_0 = 5h_c, \quad V_0 = {q_c\over 5 h_c}.
\end{equation}
We regard $\delta=h_0 T_e'(h_0)/T_e(h_0)$ as a parameter in this
section although $T_e'(h_0)$ is zero for $h_0=5h_c$ if we employ
Eq.(\ref{T_e-h}) for $T_e(h_0)$.


\subsection{Unidirectional wind}

Fig.~\ref{fig:growth_rate-uni} shows the real parts of the
eigenvalues $\omega_k^\pm(\chi)$ of the matrix $\hat A_k(\chi)$ given
by Eq.(\ref{A_k(chi)}).  They correspond to the growth rate for the
perturbation with the wave number $k$ under the unidirectional wind.
Note that the wave number $k$ is defined along the $y$ axis; 
the wave number along the crest line $k_{\rm crest}$ is given by 
\begin{equation}
k_{\rm crest} = k \cos\chi .
  \label{k_crest}
\end{equation}

The upper plots of Fig.\ref{fig:growth_rate-uni} show the dispersion
for various values of the angle $\chi$ at $(\delta, \rho)=(0,2)$,
which is in the wavy unstable regime.
The dispersion as a function $k$ 
changes significantly only for very large 
{$\tilde\chi\gtrsim 80^\circ$}, where
the validity of the present model is not so {clear}.
For the moderate value of {$\tilde\chi\lesssim 40^\circ$},
only visible difference would be the change in the wave number
(\ref{k_crest}), i.e. the wave length of the most unstable mode along
the crest line is elongated by the factor $\cos\chi$.
As we will see below, this insensitivity of the dispersions on the
slant angle $\chi$ is a common feature for all the cases we study in
the following.

In order to see the effect of non-zero $\chi$ for other values of
$(\delta,\rho)$, in the lower graphs
the dispersions of the growth rate are plotted
along the $\rho=2$ line in the wavy unstable region 
for {$\tilde\chi= 80^\circ$} together with those of
{$\tilde\chi=0$};
Even for this unrealistically large value of {$\tilde\chi=
80^\circ$}, 
the difference in the dispersions is modest.
The dispersion of the unstable modes have maxima
at finite $k$ for both {$\tilde\chi=0$ and $80^\circ$} cases in the region of
$-1< \delta < {1/2}$ with $\rho=2$ as expected from the stability
diagram in Fig.\ref{fig:nanamesouzu}.

\subsection{Transverse dune under bimodal wind}

\begin{figure}
  \centerline{
 \includegraphics[width=9cm] {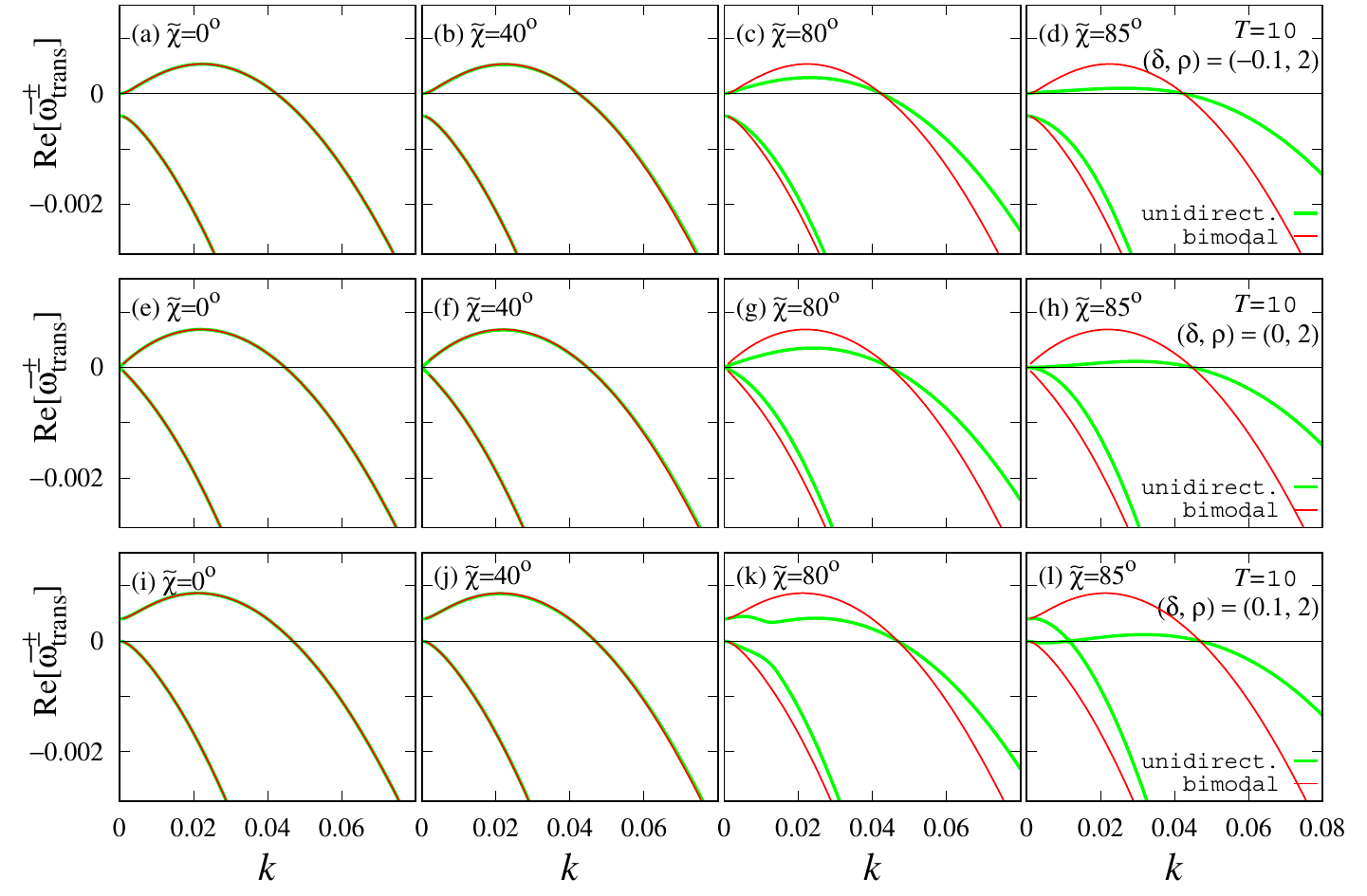}}
  \caption{The dispersions of the average growth rate
  ${\rm Re}\,[\bar\omega^\pm_{\rm trans}]$ of the transverse dune
  {under the bimodal wind with $T=10$}
   for various slant angle {$\tilde\chi$}
  in the wavy unstable regime,
  $(\delta,\rho)=(-0.1,2)$ (upper), $(0,2)$ (middle), and
  $(0.1,2)$ (lower).
  {The dispersions under the unidirectional wind are also shown
  for comparison.}
  The parameters are 
  $h_0=5$, $q=1$, $D_u=1$,
  $T_e(h_0)=1$,  $(A,B,C)=(0.1,0.8,0.2)$
  with $T_e'(h_0)=(T_e(h_0)/h_0) \delta$ and $D_d=D_u\rho$.
  }
  \label{fig:growth_rate-bimod-trans}
\end{figure}

Fig.\ref{fig:growth_rate-bimod-trans} shows the dispersions of the
average growth rate Eq.(\ref{ave_omega_trans}) 
for the transverse dune under the bimodal wind for various slant angle
{$\tilde\chi$}; 
the growth rates under the unidirectional wind,
${\rm Re}\,[\omega_k^\pm(\chi)]$,
for the same slant angle {$\tilde\chi$} are also shown for comparison.
The wind duration of each direction $T$ is 10 in the unit of
Eq.(\ref{unit_system}) and the model parameters are taken
in the wavy unstable regime for the unidirectional wind, i.e.
$(\delta,\rho)=(-0.1,2)$, (0,2), and (0.1,2), thus
the wave number for the most unstable mode $k_{\rm max}$ is non-zero.

As a function of $k$, the dispersion for the bimodal wind (the red
curves) hardly depends on the slant angle even for the very large
{$\tilde\chi\gtrsim 80^\circ$}, where the dispersions for the
unidirectional wind (the green curves) deviates significantly from that
of {$\tilde\chi=0$}.  
This can be understood as follows; for small $T$
which satisfies ${\omega}^\pm_k(\chi)T\ll 1$, the evolution
matrix $\hat\Lambda_{\rm trans}$ given by Eq.(\ref{Lambda_trans}) 
can be approximated as
\begin{equation}
\hat\Lambda_{\rm trans} \approx
  1+ \hat A^*_k(\chi) T +  \hat A_k(\chi) T  
  = 1 +   \hat A_k(0) 2 T  ,
  \label{Lambda_trans-approx}
\end{equation}
thus
\begin{equation}
  {\bar\omega}_{\rm trans}^\pm(k;\chi) \approx \omega_k^\pm(0).
\end{equation}
This means that the effects of the slant angle from the two winds with
$\chi$ and $-\chi$ in Fig.~\ref{fig:bimodal-wind} cancel each
other after one period of the bimodal wind\footnote{
It is interesting to see that the average growth rate for the
bimodal wind can be larger than the growth rate for the unidirectional
wind.  This means that the largest absolute value of the eigenvalue of
the evolution matrix $\hat\Lambda_{\rm trans}$ is larger than the
product of the largest absolute values of the eigenvalue of each
factor matrix in Eq.(\ref{Lambda_trans}).  
}.

Fig.\ref{fig:growth_rate-bimod-trans-T} shows 
the duration time $T$ dependence of the dispersions for
{$\tilde\chi=80^\circ$} and $(\delta,\rho)=(0.1,2)$.
For such a large value of slant angle,
the dispersion for the bimodal wind with $T=10$ is still close to that
of {$\tilde\chi=0$}, but
significantly different from that for the unidirectional wind with the
same value of {$\tilde\chi=80^\circ$}.
For very large $T$,
one can see that the dispersions for the bimodal wind become closer to
those for the unidirectional wind as one should expect; the bimodal
wind with very long $T$ should look like the unidirectional wind.  
We do not show the $T$-dependence for
a smaller slant angle, e.g. {$\tilde\chi\lesssim 40^\circ$}, but
the dispersions hardly depend on the duration of each wind $T$
because the dispersions for the bimodal wind and the unidirectional
wind are already very close to each other and to those for the
{$\tilde\chi=0$} case.

\begin{figure}[t]
\centerline{
\includegraphics[width=8.9cm] {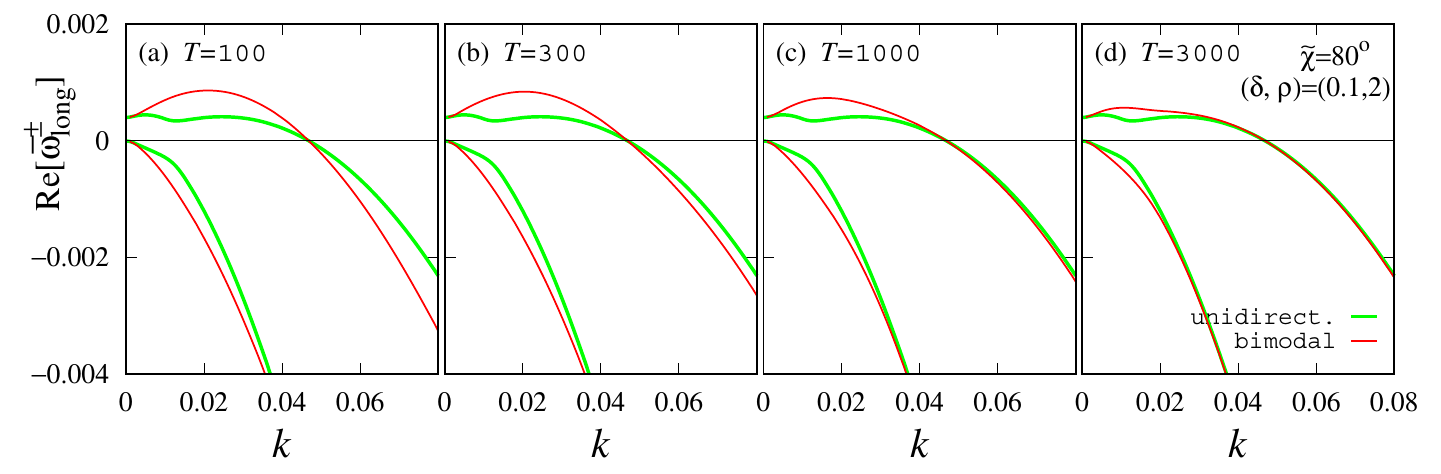}}
\caption{The dispersions of the average growth rate
${\rm Re}\,[\bar\omega^\pm_{\rm trans}]$ of the transverse dune
  under the bimodal wind
  for various $T$ with {$\tilde\chi=80^\circ$} at
  $(\delta,\rho)=(0.1,2)$
  {along with those under the unidirectional wind for comparison.}
 The parameters are $h_0=5$, $q=1$,
  $D_u=1$, $T_e(h_0)=1$,  $(A,B,C)=(0.1,0.8,0.2)$ with
  $T_e'(h_0)=(T_e(h_0)/h_0) \delta$ and $D_d=D_u\rho$.
}
  \label{fig:growth_rate-bimod-trans-T}
\end{figure}

\subsection{Longitudinal dune under bimodal wind}

Fig.\ref{fig:growth_rate-long} shows the dispersions of the average
growth rate given by Eq.(\ref{ave_omega_long}) of the longitudinal
dune under the bimodal wind for various slant angle {$\tilde\chi$}
with $T=10$ and $(\delta,\rho)=(-0.1,2)$, (0,2), and (0.1,2).
It is remarkable that
the wave number for the most unstable mode $k_{\rm max}$ is
always zero for the longitudinal dune under the bimodal wind,
even though the parameters $(\delta, \rho)$ are in the wavy unstable
regime for the unidirectional wind.

This can be understood as follows; for ${\omega}^\pm_k(\chi)T\ll 1$,
the evolution matrix $\hat\Lambda_{\rm long}$ of
Eq.(\ref{Lambda_long}) can be approximated as
\begin{widetext}
\begin{equation}
  \hat\Lambda_{\rm long} \approx
  1 + \hat\sigma_z\hat A_k(\chi)\hat\sigma_z T + \hat A_k(\chi) T
  =
  1 + \hat A_k^{\rm diag}(\chi) 2T,
  \label{Lambda_long-approx}
\end{equation}
where, $\hat A_k^{\rm diag}(\chi)$ represents the diagonal part
of $\hat A_k(\chi)$,
\begin{align}
   \hat A_k^{\rm diag}(\chi)  \equiv 
   \left(\begin{array}{cc}
    -(BD_d+CD_u)k^2, & 0 \\
    0, &
    AqT_e'/h_0+A(D_d-D_u)/h_0\,ik\tan\chi - (CD_d+BD_u)k^2
  \end{array}\right) .
  \label{A_k(chi)|diag}
\end{align}
Thus we have
\begin{equation}
  \bar\omega_{\rm long}^\pm(k;\chi) \approx
  \left\{\begin{array} {l}
-(BD_d+CD_u)k^2 \\
AqT_e'/h_0+A(D_d-D_u)/h_0\,ik\tan\chi - (CD_d+BD_u)k^2
  \end{array} \right. ,
\end{equation}
\end{widetext}
where $\bar\omega_{\rm long}^+(k;\chi)$ 
corresponds to the one with larger real part in RHS 
and $\bar\omega_{\rm long}^-(k;\chi)$ to the other.
{From this expression for $\bar\omega_{\rm long}^\pm$, we can
understand the reason why the two modes almost degenerate in the plots
of the middle row of Fig.\ref{fig:growth_rate-long}, where we set
$B=C=0.5$ and $T_e' =0$.}

\begin{figure}[b]
  \centering
  {\includegraphics[width=9cm]
  {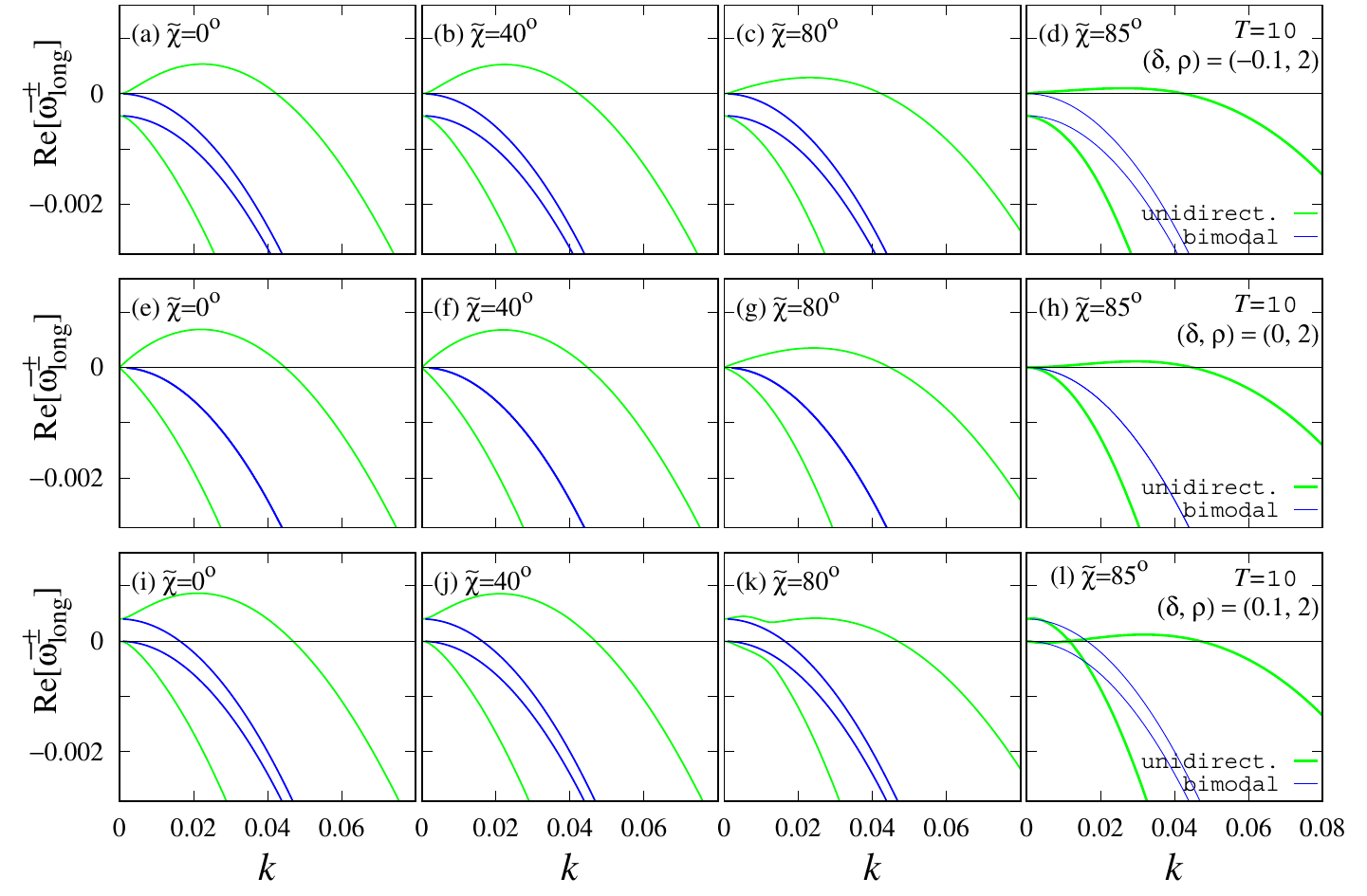}}
  \caption{The dispersions of the average growth rate
  ${\rm Re}\,[\bar\omega^\pm_{\rm long}]$ of the longitudinal dune
  {under the bimodal wind}
  with $T=10$ for various slant angle {$\tilde\chi$}
  in the wavy unstable regime,
  $(\delta,\rho)=(-0.1,2)$ (upper), $(0,2)$ (middle), and
  $(0.1,2)$ (lower).
  {The dispersions under the unidirectional wind are also shown
  for comparison.}
   The parameters are 
  $h_0=5$, $q=1$, $D_u=1$,
  $T_e(h_0)=1$,  $(A,B,C)=(0.1,0.5,0.5)$
 with $T_e'(h_0)=(T_e(h_0)/h_0) \delta$ and $D_d=D_u\rho$.
}
  \label{fig:growth_rate-long}
\end{figure}
\begin{figure}[b]
  \centerline{\includegraphics[width=9cm]
  {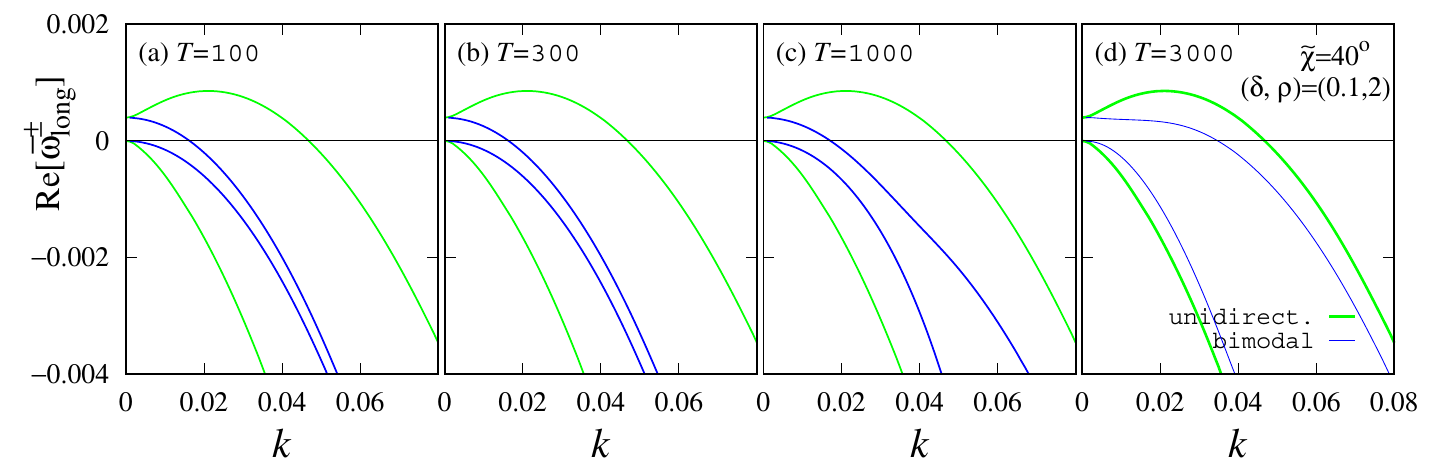}}
  \caption{The dispersions of the average growth rate
  ${\rm Re}\,[\bar\omega^\pm_{\rm long}]$ of the longitudinal dune
  {under the bimodal wind}
  for various values of $T$ with the slant angle
  {$\tilde\chi=40^\circ$}
  as $(\delta,\rho)=(0.1,2)$
  {along with those under the unidirectional wind for
  comparison.}  
  The parameters are $h_0=5$, $q=1$, $D_u=1$,
  $T_e(h_0)=1$,  $(A,B,C)=(0.1,0.5,0.5)$
  with $T_e'(h_0)=(T_e(h_0)/h_0) \delta$ and $D_d=D_u\rho$.
}
  \label{fig:growth_rate-long-T}
\end{figure}

The duration time $T$ dependence of the dispersion is shown in
Fig.~\ref{fig:growth_rate-long-T};
the dispersion for the bimodal wind approaches toward those for the
unidirectional wind for very large $T$ as is expected.

\section{Simulations under bimodal winds of isolated straight
transverse and longitudinal dunes with finite length}\label{Sec:numerics}
\begin{figure*}
\centerline{\includegraphics[width=160mm]{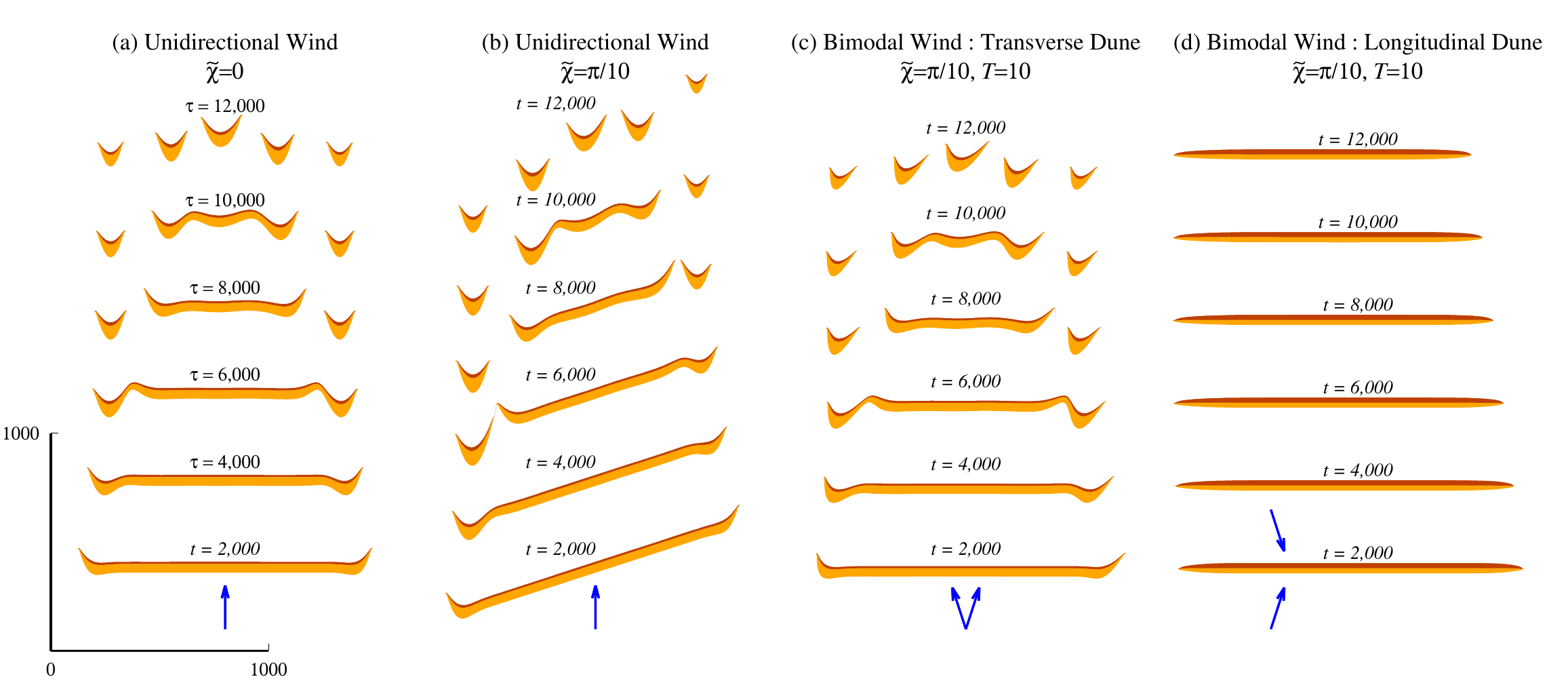}}
\caption{The time evolutions of the dune obtained by simulations of
  the crest line model for the unidirectional wind with
  {$\tilde\chi=0$} (a)
  and $\pi/10$ (b), the transverse dune (c) and the longitudinal dune
  (d) under the bimodal wind dune with {$\tilde\chi=\pi/10$} and
  $T=10$.  The arrows show the directions of the wind for each case.  
  The parameters are $(\delta,\rho)=(0,2)$, which is in the wavy
  unstable regime for the perturbation.  The geometrical
  {parameters} for the cross section are $(A,B,C)=(0.1,0.8,0.2)$
  for the dune under the unidirectional wind and the transverse dune,
  $(A,B,C)=(0.1,0.5,0.5)$ for the longitudinal dune.  The initial
  state are given in the text.
  The advances of the dunes are represented by the vertical positions
  except for the case of the longitudinal dune, in which case 
  the dune does not actually advance in the vertical direction.
  }
\label{fig:Simulation}
\end{figure*}

We perform numerical simulations for the crest line model (\ref{eq:h}) and
(\ref{eq:x})
for an isolated straight dune with finite length, and
compare them with the results of linear analysis given above.
The initial state is given by
\begin{align}
  h(y) & = \left\{\begin{array}{ll}
    h_0 (L_y/2+y)/L_b & \mbox{for $-L_y/2\le y \le -L_y/2+ L_b$} \\
  h_0 &\hspace*{-1.5em} \mbox{for $-L_y/2+L_b< y < L_y/2-L_b$} \\
    h_0\, (L_y/2-y)/L_b & \mbox{for $L_y/2-L_b \le y \le L_y/2$} \\
    0  & \mbox{otherwise}
  \end{array}\right. ,
  \\
  x(y) & = y \tan\chi
\end{align}
with the 1\% of noise added to $h$.
Here, $L_y=1500$, $L_b=L_y/20$, and $h_0=5$ in the units of Eq.(\ref{unit}).
We take $D_d=2$, thus
for the initial state, $(\delta,\rho) =(0, 2)$, it is in the wavy
unstable regime with $k_{\rm max}\approx {0.22\times 10^{-1}}$ and 
$\omega^+_{k_{\rm max}}\approx {0.69\times 10^{-3}}$ for $\chi=0$.

Since the model is defined in the way that the wind always blows in
the $x$ direction, the crest line $\big(x(y),h(y)\big)$ needs to be
rotated in the $x-y$ plane every time when the wind direction changes.
The difficulty in this procedure is that the function $x(y)$, which
represents the position of the crest line, may become multi-valued
after the rotation, to which case the present model is not extended.
To avoid this difficulty, we employ simplified transformations 
instead of the rotation, namely,
\begin{equation}
  \Big(x(y),\; h(y)\Big) \quad\rightarrow\quad 
  \Big(x(y)-2y\tan\chi,\; h(y)\Big),
\end{equation}
for the bimodal wind in the case of transverse dune,
\begin{equation}
  \Big(x(y),\; h(y)\Big) \quad\rightarrow\quad 
  \Big(-x(-y)-2y\tan\chi,\; h(-y)\Big),
\end{equation}
in the case of longitudinal dune every time when the wind direction
changes.  
These are analogous to the ones used in the linear analysis
and the difference between this simplified transformations and
the rotations is small when $\chi$ and $x(y)-y\tan\chi$ are small.

Fig.\ref{fig:Simulation} shows the time evolutions of sand dune for
{$\tilde\chi=0$} (a) and $\pi/10$ (b) under the unidirectional
wind, and those for the transverse (c) and the longitudinal (d) dunes
under the bimodal wind with $T=10$ and {$\tilde\chi=\pi/10$}.
The time evolution of the dune under the unidirectional wind with
{$\tilde\chi=\pi/10$} (b) and that of the transverse dune under
the bimodal wind (c) are quite similar to that of the one under the
unidirectional wind with {$\tilde\chi=0$} (a); all of them show
the instability with nearly the same wave number.  On the other hand,
in the case of the longitudinal dune with {$\tilde\chi=\pi/10$}
under the bimodal wind, the instability does not appear and the
straight crest line is stable even though the parameters are in the
region of the wavy unstable regime for the unidirectional wind.  These
results are completely consistent with
those of the linear analysis given in Sec.\ref{numerical_growth}.

{
  One may notice that the dune length decreases in time in our
simulations while it has been observed that dunes often elongate in
the average wind direction \cite{Tsoar-2004}.  The reason for the
shortening of the dune length in our simulations is that our
simulations are performed under the condition of no influx, i.e.
$f_{\rm in}=0$, thus the system lose sand grains at the ends of the
dune, where the efficiency rate $T_e<1$ because $h<h_c$.  It should be
noted that, even with this setting, we observe the cases where the
dune elongate toward the downwind direction if we use larger skew
angle.
}

\section{Concluding remarks}

\paragraph{{Advantages and disadvantages of the crest line model:}}

{
The crest line model used in the present study is one of the simplest
model for the dune dynamics, where the dune configuration is described
only by the position $x$ and the height $h$ of the crest as a function
of $y$, assuming a similar triangle for the dune cross section along
the wind direction.  Its dynamics is determined by the sand
transportations along the longitudinal and the lateral directions,
which are characterized by a small number of phenomenological
parameters and functions, i.e.  $f_{\rm in}$, $q(h)$, $T_e(h)$, $D_u$,
and $D_d$.  Despite its simplicity, the model is capable to reproduce
basic features of the dune dynamics.
}

{
Due to its simplicity, there are some obvious drawbacks.  
The model cannot describe the evolution of dunes of general forms,
such as star dunes, that cannot be characterized by the geometrical
parameters.
%
%
The model
in the original form is for a single dune, thus ignores the
interaction between dunes.
The parameters are largely phenomenological, thus it is not
straightforward to derive them from elementary processes;
the lateral sand fluxes, for example,  are characterized simply by
the diffusion constants $D_u$ and $D_d$, but
the diffusion processes
actually represent combined effects of biased random motion of sand
grains due to saltation, reptation, creeping, avalanche, etc.
These processes are excited by turbulent and helical flows over the
dune, thus $D_u$ and $D_d$ may well depend on $q$ or some other
parameters/variables although they are assumed to be constant in the
present paper.
}

{
The advantages of the crest line model are also due to its simplicity;
we can perform the stability analysis almost by hand even for the
cases under the bimodal wind, and can analyze the results rather
systematically to understand how the longitudinal dune becomes stable
under the bimodal wind with large variation of the wind direction.
}

{
The present analysis show that the developed straight longitudinal
dune with $\delta=0$, or $T_e'=0$, is stable. However, it is often
observed in the field that the longitudinal dunes especially without
vegetation are undulating and are called {seif\cite{Pye-2009}}. 
We do not know yet how
the vegetation effects may be incorporated in the crest line model,
but our results suggest that the undulation in the longitudinal dune
may arise during the dune developing process because the dispersions
in the lower panel of Fig.\ref{fig:growth_rate-long} for
$(\delta,\rho)=(0.1,2)$ have an unstable part around $k=0$; 
The dune with $\delta >0$ is growing one because it is not yet
saturated, i.e. $T_e'>0$.
These unstable modes can result in
the undulation during the growth process if there are some effects
that suppress the instability of the uniform mode with $k=0$, thus the
maximum growth mode becomes at $k\ne 0$.
}

\paragraph{{Parameter estimate:}}
{
Let us now estimate some of the parameter values in our model.
Important ones are the critical dune height $h_c$ and the saturated
sand flux $q_c$ over the crest of the critical height $h_c$.  From the
numerical results of fluid dynamics
modelling\cite{Warren,Momiji-2001}, they might be estimated as%
\footnote{{
  From the numerical estimates of the trapping efficiency $T_e$
(Fig.3.3 at p.50 of \cite{Warren}) and the friction velocity $v^*$
(Fig.4.5 at p.88 of \cite{Momiji-2001}), we obtain  the estimates
$h_c\sim 5$ and $u^*\sim$ 0.5 m/s.  For this value of the friction
velocity, the mass flux at the crest $q_{\rm mass}$ can be estimated
as $q_{\rm mass}\sim 0.01$ kg/m s from Fig.1.8 in p.28 of
\cite{Warren} .  The volume flux $q_c$ is given by $q_c=q_{\rm
mass}/\rho$ with the mass density of sand 
$\rho\approx 1.7\times 10^{3} {\rm\, kg\, m^{-3}}$.}}
\begin{equation}
  h_c \sim 5 {\rm\, m}, \qquad
  q_c \sim 5.9 \times 10^{-6} {\rm\, m^2/s}.
  \label{est-hc-qc}
\end{equation}
This estimate for $q_c$ leads to the migration speed of
dune $v$ of the crest height $h_c$ as
\begin{equation}
  v={q_c\over h_c} \sim 38 {\rm\, m/year}
\end{equation}
in the case of perfect trapping without influx/outflux, i.e.
$T_e=1$ and $f_{\rm in}=0$.  This is actually consistent with the
field observation data for the migration speed $v_{\rm obs}$ 
of the dunes with the height of 5 m
(Fig.23.24 of \cite{Cooke}),
\begin{equation}
  15 {\rm\, m/year}\lesssim v_{\rm obs}\lesssim 50 {\rm\, m/year}.
\end{equation} }

{
For the time unit of Eq.(\ref{unit}),
the above estimates of  Eq.(\ref{est-hc-qc}) gives 
\begin{equation}
  {h_c^2\over q_c} \sim 0.13 {\rm\; year} .
\end{equation}
Thus, our choice of $T=10$ for the duration of bimodal wind
corresponds to 1.3 year, which is a little too longer as it should be
0.5 year, but we believe that the discrepancy is not significant,
considering the accuracy of the above estimate.  It should be also
pointed out that most of our numerical results would barely change
even if we use smaller value for $T$ because $T=10$ is
already well in the small $T$ limit, and the approximations in
Eqs.(\ref{Lambda_trans-approx}) and (\ref{Lambda_long-approx}) are
very good.
}

{
The spatial scale in the $y$ direction, or the lateral
direction, is given by  $h_c\sqrt{D_u/q_c}$ in Eq.(\ref{unit}). 
This contains  a
phenomenological parameter $D_u$, whose value should be a
result of various processes under the influence of fluctuating wind
averaged over the duration time, therefore, difficult to estimate
from ``microscopic processes''.  However, 
the present stability analysis would give an
estimate for $D_u$ as follows.
For the transverse dune under the unidirectional wind,
the dispersion curves in Fig.\ref{fig:growth_rate-uni} for
$\tilde\chi=0^\circ$ and $(\delta,\rho)=(0, 2)$,
which corresponds to the well-developed dune,
show the most unstable
mode  at around $k_{\rm max}\approx 0.025$, 
or in the length scale, it is 
$\lambda_{\rm max}\approx 250$ m.  This actually
coincides with the lateral size of the barchans with the
height $h_0=5$ in Fig.\ref{fig:Simulation}(a).  If we know observation
data for the  barchan width $W_{\rm obs}$, then we can estimate the
value of $D_u$ from the relation
\begin{equation}
  W_{\rm obs} \approx \lambda_{\rm max} h_c \sqrt{D_u\over q_c}.
\end{equation}}

{
As a rough estimate, if we use $W_{\rm obs}\approx 250$ m for the dune
with the height $h_0=5h_c\approx 25$ m, then 
the unit length in the $y$ direction is estimated as
\begin{equation}
  h_c\sqrt{D_u\over q_u} \sim 1 {\rm\, m},
  \label{es-y_unit}
\end{equation}
which means that the length scale in the $y$ direction, i.e. the
direction perpendicular to the wind,  should be 
reduced by the factor 1/5 in Fig.\ref{fig:Simulation}.
This also gives
\begin{equation}
  D_u \sim {q_c\over 25} \sim 0.24 \times 10^{-6} {\rm\, m^2/s},
  \label{eq-Du}
\end{equation}
which turns out to be several orders of magnitude smaller than 4.0
$\rm m^2/s$, i.e. the value used for the diffusion constant by Schw\"ammle
and Herrmann\cite{Schwammle-2005}, but it should be pointed out that
$D_u$ in the crest line model is a phenomenological parameter, thus
cannot be compared with the ``microscopic'' diffusion constant
directly.}
{
  With the estimate of Eq.(\ref{eq-Du}), 
  the relationship between $\chi$ and $\tilde\chi$
given by Eq.(\ref{tilde_chi}) becomes 
\begin{equation}
  \tan\chi \approx 5\tan\tilde\chi,
  \label{chi-tilde_chi}
\end{equation}
which means $\chi > \tilde\chi$.
Since most of our numerical results for $\tilde\chi=40^\circ$ are very
close to the ones for $\tilde\chi=0^\circ$, 
we can conclude that the slant angle effect is very small for any
value of $\chi<45^\circ$ except for the fact the spatial modulation
along the crest line is elongated as Eq.(\ref{k_crest}).
}


\paragraph{Summary:}
Using the crest line model, we examined the dynamical stability of the
isolated straight sand dune under the unidirectional and the bimodal
winds with the non-zero slant angle $\chi$.  In the case of the
unidirectional wind, we found that the linear stability for $\chi\ne
0$ remains the
same with that for the wind with  $\chi=0$.
We extended the linear stability analysis to the case of bimodal winds
and obtained the expressions for the growth rates for perturbation.
Assuming the wind strengths and durations are the same for {the
both directions of} the bimodal
winds, we examined the two cases: the transverse
dune, whose crest line is perpendicular to the average wind direction,
and the longitudinal dune, whose crest line is parallel to the average
wind.
{
For the case of the unidirectional wind with
$\chi\ne 0$ and for the case of the transverse dune,
the dispersion curves of the growth rates}
virtually do not depend on the slant
angle $\chi$ and are almost the same as those for the unidirectional
wind for $\chi=0$ except for the case with very large slant angle,
i.e. {$\tilde\chi\gtrsim 80^\circ$}, where the validity of the
present model is not {clear}.  
On the other hand, in the case of the longitudinal dune, the
dispersion curves of the growth rate change from those for the
unidirectional wind so that the largest growth rate is always at $k=0$
even when the parameters are in the wavy unstable regime for the
unidirectional wind case.
This may be interpreted as a result of the fact that
the growth of the perturbation with $k\ne 0$ is canceled by the
alternating wind from opposite sides of the crest line
even though it grows during each duration period of the bimodal wind. 
Mathematically, this cancellation is due to the existence of the
decaying mode in the unidirectional case;  The stabilization occurs
because a major part of the growing mode is converted to the decaying
mode whenever the wind direction switches. 
%

{
The present study provides a theoretical framework for the stability
analysis of sand dunes under the simple bimodal winds,  and we
demonstrated the stabilizing mechanism of a straight crest line for the
longitudinal dune.   This result qualitatively corresponds to the
field observations and the computer simulations that the straight
dunes tend to align to the average wind direction in the case of large
seasonal variation of wind direction.
It should be noted, however, that unvegetated linear dunes are often
observed to have an undulating crest line, and straight dunes are
usually vegetated\cite{Tsoar, Parteli-2009}.  It is also intriguing
that almost straight longitudinal dune-like structures have been found
in the images of Titan\cite{Lorenz-2006,tokano-2008}.  
Understanding detailed conditions
for the undulation to appear/disappear in the longitudinal dunes is a
theoretical problem in the next step.
}

\bibliography{references}

\begin{thebibliography}{37}%
\makeatletter
\providecommand \@ifxundefined [1]{%
 \@ifx{#1\undefined}
}%
\providecommand \@ifnum [1]{%
 \ifnum #1\expandafter \@firstoftwo
 \else \expandafter \@secondoftwo
 \fi
}%
\providecommand \@ifx [1]{%
 \ifx #1\expandafter \@firstoftwo
 \else \expandafter \@secondoftwo
 \fi
}%
\providecommand \natexlab [1]{#1}%
\providecommand \enquote  [1]{``#1''}%
\providecommand \bibnamefont  [1]{#1}%
\providecommand \bibfnamefont [1]{#1}%
\providecommand \citenamefont [1]{#1}%
\providecommand \href@noop [0]{\@secondoftwo}%
\providecommand \href [0]{\begingroup \@sanitize@url \@href}%
\providecommand \@href[1]{\@@startlink{#1}\@@href}%
\providecommand \@@href[1]{\endgroup#1\@@endlink}%
\providecommand \@sanitize@url [0]{\catcode `\\12\catcode `\$12\catcode
  `\&12\catcode `\#12\catcode `\^12\catcode `\_12\catcode `\%12\relax}%
\providecommand \@@startlink[1]{}%
\providecommand \@@endlink[0]{}%
\providecommand \url  [0]{\begingroup\@sanitize@url \@url }%
\providecommand \@url [1]{\endgroup\@href {#1}{\urlprefix }}%
\providecommand \urlprefix  [0]{URL }%
\providecommand \Eprint [0]{\href }%
\providecommand \doibase [0]{http://dx.doi.org/}%
\providecommand \selectlanguage [0]{\@gobble}%
\providecommand \bibinfo  [0]{\@secondoftwo}%
\providecommand \bibfield  [0]{\@secondoftwo}%
\providecommand \translation [1]{[#1]}%
\providecommand \BibitemOpen [0]{}%
\providecommand \bibitemStop [0]{}%
\providecommand \bibitemNoStop [0]{.\EOS\space}%
\providecommand \EOS [0]{\spacefactor3000\relax}%
\providecommand \BibitemShut  [1]{\csname bibitem#1\endcsname}%
\let\auto@bib@innerbib\@empty
\bibitem [{\citenamefont {Bagnold}(1941)}]{Bagnold}%
  \BibitemOpen
  \bibfield  {author} {\bibinfo {author} {\bibfnamefont {R.}~\bibnamefont
  {Bagnold}},\ }\href@noop {} {\emph {\bibinfo {title} {The Physics of Blown
  Sand and Desert Dunes}}}\ (\bibinfo  {publisher} {Methuen},\ \bibinfo
  {address} {London},\ \bibinfo {year} {1941})\BibitemShut {NoStop}%
\bibitem [{\citenamefont {Pye}\ and\ \citenamefont {Tsoar}(1990)}]{Tsoar}%
  \BibitemOpen
  \bibfield  {author} {\bibinfo {author} {\bibfnamefont {K.}~\bibnamefont
  {Pye}}\ and\ \bibinfo {author} {\bibfnamefont {H.}~\bibnamefont {Tsoar}},\
  }\href@noop {} {\emph {\bibinfo {title} {Aeolian Sand and Sand Dunes}}}\
  (\bibinfo  {publisher} {Unwin Hyman},\ \bibinfo {address} {London},\ \bibinfo
  {year} {1990})\BibitemShut {NoStop}%
\bibitem [{\citenamefont {Cooke}\ \emph {et~al.}(1993)\citenamefont {Cooke},
  \citenamefont {Warren},\ and\ \citenamefont {Goudie}}]{Cooke}%
  \BibitemOpen
  \bibfield  {author} {\bibinfo {author} {\bibfnamefont {R.~U.}\ \bibnamefont
  {Cooke}}, \bibinfo {author} {\bibfnamefont {A.}~\bibnamefont {Warren}}, \
  and\ \bibinfo {author} {\bibfnamefont {A.~S.}\ \bibnamefont {Goudie}},\
  }\href@noop {} {\emph {\bibinfo {title} {Desert Geomorphology}}}\ (\bibinfo
  {publisher} {UCL Press},\ \bibinfo {address} {London},\ \bibinfo {year}
  {1993})\BibitemShut {NoStop}%
\bibitem [{\citenamefont {Lancaster}(1995)}]{Lancaster}%
  \BibitemOpen
  \bibfield  {author} {\bibinfo {author} {\bibfnamefont {N.}~\bibnamefont
  {Lancaster}},\ }\href@noop {} {\emph {\bibinfo {title} {Geomorphology of
  Desert Dunes}}}\ (\bibinfo  {publisher} {Routledge},\ \bibinfo {address}
  {London},\ \bibinfo {year} {1995})\BibitemShut {NoStop}%
\bibitem [{\citenamefont {Warren}(2013)}]{Warren}%
  \BibitemOpen
  \bibfield  {author} {\bibinfo {author} {\bibfnamefont {A.}~\bibnamefont
  {Warren}},\ }\href@noop {} {\emph {\bibinfo {title} {Dunes: dynamics,
  morphology, history}}}\ (\bibinfo  {publisher} {Wiley \& Sons},\ \bibinfo
  {address} {UK},\ \bibinfo {year} {2013})\BibitemShut {NoStop}%
\bibitem [{\citenamefont {Telfer}\ \emph {et~al.}(2018)\citenamefont {Telfer},
  \citenamefont {Parteli}, \citenamefont {Radebaugh}, \citenamefont {Beyer},
  \citenamefont {Bertrand}, \citenamefont {Forget}, \citenamefont {Nimmo},
  \citenamefont {Grundy}, \citenamefont {Moore}, \citenamefont {Stern},
  \citenamefont {Spencer}, \citenamefont {Lauer}, \citenamefont {Earle},
  \citenamefont {Binzel}, \citenamefont {Weaver}, \citenamefont {Olkin},
  \citenamefont {Young}, \citenamefont {Ennico},\ and\ \citenamefont
  {Runyon}}]{Telfer-2018}%
  \BibitemOpen
  \bibfield  {author} {\bibinfo {author} {\bibfnamefont {M.~W.}\ \bibnamefont
  {Telfer}}, \bibinfo {author} {\bibfnamefont {E.~J.~R.}\ \bibnamefont
  {Parteli}}, \bibinfo {author} {\bibfnamefont {J.}~\bibnamefont {Radebaugh}},
  \bibinfo {author} {\bibfnamefont {R.~A.}\ \bibnamefont {Beyer}}, \bibinfo
  {author} {\bibfnamefont {T.}~\bibnamefont {Bertrand}}, \bibinfo {author}
  {\bibfnamefont {F.}~\bibnamefont {Forget}}, \bibinfo {author} {\bibfnamefont
  {F.}~\bibnamefont {Nimmo}}, \bibinfo {author} {\bibfnamefont {W.~M.}\
  \bibnamefont {Grundy}}, \bibinfo {author} {\bibfnamefont {J.~M.}\
  \bibnamefont {Moore}}, \bibinfo {author} {\bibfnamefont {S.~A.}\ \bibnamefont
  {Stern}}, \bibinfo {author} {\bibfnamefont {J.}~\bibnamefont {Spencer}},
  \bibinfo {author} {\bibfnamefont {T.~R.}\ \bibnamefont {Lauer}}, \bibinfo
  {author} {\bibfnamefont {A.~M.}\ \bibnamefont {Earle}}, \bibinfo {author}
  {\bibfnamefont {R.~P.}\ \bibnamefont {Binzel}}, \bibinfo {author}
  {\bibfnamefont {H.~A.}\ \bibnamefont {Weaver}}, \bibinfo {author}
  {\bibfnamefont {C.~B.}\ \bibnamefont {Olkin}}, \bibinfo {author}
  {\bibfnamefont {L.~A.}\ \bibnamefont {Young}}, \bibinfo {author}
  {\bibfnamefont {K.}~\bibnamefont {Ennico}}, \ and\ \bibinfo {author}
  {\bibfnamefont {K.}~\bibnamefont {Runyon}},\ }\href {DOI:
  10.1126/science.aao2975} {\bibfield  {journal} {\bibinfo  {journal}
  {Science}\ }\textbf {\bibinfo {volume} {360}},\ \bibinfo {pages} {992}
  (\bibinfo {year} {2018})}\BibitemShut {NoStop}%
\bibitem [{\citenamefont {Andreotti}\ \emph
  {et~al.}(2002{\natexlab{a}})\citenamefont {Andreotti}, \citenamefont
  {Claudin},\ and\ \citenamefont {Douady}}]{Andreotti-2002a}%
  \BibitemOpen
  \bibfield  {author} {\bibinfo {author} {\bibfnamefont {B.}~\bibnamefont
  {Andreotti}}, \bibinfo {author} {\bibfnamefont {P.}~\bibnamefont {Claudin}},
  \ and\ \bibinfo {author} {\bibfnamefont {S.}~\bibnamefont {Douady}},\ }\href
  {https://doi.org/10.1140/epjb/e2002-00236-4} {\bibfield  {journal} {\bibinfo
  {journal} {Eur. Phys. J. B}\ }\textbf {\bibinfo {volume} {28}},\ \bibinfo
  {pages} {321} (\bibinfo {year} {2002}{\natexlab{a}})}\BibitemShut {NoStop}%
\bibitem [{\citenamefont {Andreotti}\ \emph
  {et~al.}(2002{\natexlab{b}})\citenamefont {Andreotti}, \citenamefont
  {Claudin},\ and\ \citenamefont {Douady}}]{Andreotti-2002b}%
  \BibitemOpen
  \bibfield  {author} {\bibinfo {author} {\bibfnamefont {B.}~\bibnamefont
  {Andreotti}}, \bibinfo {author} {\bibfnamefont {P.}~\bibnamefont {Claudin}},
  \ and\ \bibinfo {author} {\bibfnamefont {S.}~\bibnamefont {Douady}},\ }\href
  {https://doi.org/10.1140/epjb/e2002-00237-3} {\bibfield  {journal} {\bibinfo
  {journal} {Eur. Phys. J. B}\ }\textbf {\bibinfo {volume} {28}},\ \bibinfo
  {pages} {341} (\bibinfo {year} {2002}{\natexlab{b}})}\BibitemShut {NoStop}%
\bibitem [{\citenamefont {Kroy}\ \emph
  {et~al.}(2002{\natexlab{a}})\citenamefont {Kroy}, \citenamefont {Sauermann},\
  and\ \citenamefont {Herrmann}}]{Kroy-2002a}%
  \BibitemOpen
  \bibfield  {author} {\bibinfo {author} {\bibfnamefont {K.}~\bibnamefont
  {Kroy}}, \bibinfo {author} {\bibfnamefont {G.}~\bibnamefont {Sauermann}}, \
  and\ \bibinfo {author} {\bibfnamefont {H.~J.}\ \bibnamefont {Herrmann}},\
  }\href {\doibase 10.1103/PhysRevLett.88.054301} {\bibfield  {journal}
  {\bibinfo  {journal} {Phys. Rev. Lett.}\ }\textbf {\bibinfo {volume} {88}},\
  \bibinfo {pages} {054301} (\bibinfo {year} {2002}{\natexlab{a}})}\BibitemShut
  {NoStop}%
\bibitem [{\citenamefont {Kroy}\ \emph
  {et~al.}(2002{\natexlab{b}})\citenamefont {Kroy}, \citenamefont {Sauermann},\
  and\ \citenamefont {Herrmann}}]{Kroy-2002b}%
  \BibitemOpen
  \bibfield  {author} {\bibinfo {author} {\bibfnamefont {K.}~\bibnamefont
  {Kroy}}, \bibinfo {author} {\bibfnamefont {G.}~\bibnamefont {Sauermann}}, \
  and\ \bibinfo {author} {\bibfnamefont {H.~J.}\ \bibnamefont {Herrmann}},\
  }\href {\doibase 10.1103/PhysRevE.66.031302} {\bibfield  {journal} {\bibinfo
  {journal} {Phys. Rev. E}\ }\textbf {\bibinfo {volume} {66}},\ \bibinfo
  {pages} {031302} (\bibinfo {year} {2002}{\natexlab{b}})}\BibitemShut
  {NoStop}%
\bibitem [{\citenamefont {Sauermann}\ \emph {et~al.}(2001)\citenamefont
  {Sauermann}, \citenamefont {Kroy},\ and\ \citenamefont
  {Herrmann}}]{Sauermann-2001}%
  \BibitemOpen
  \bibfield  {author} {\bibinfo {author} {\bibfnamefont {G.}~\bibnamefont
  {Sauermann}}, \bibinfo {author} {\bibfnamefont {K.}~\bibnamefont {Kroy}}, \
  and\ \bibinfo {author} {\bibfnamefont {H.~J.}\ \bibnamefont {Herrmann}},\
  }\href {\doibase 10.1103/PhysRevE.64.031305} {\bibfield  {journal} {\bibinfo
  {journal} {Phys. Rev. E}\ }\textbf {\bibinfo {volume} {64}},\ \bibinfo
  {pages} {031305} (\bibinfo {year} {2001})}\BibitemShut {NoStop}%
\bibitem [{\citenamefont {Courrech~du Pont}\ \emph {et~al.}(2014)\citenamefont
  {Courrech~du Pont}, \citenamefont {Narteau},\ and\ \citenamefont
  {Gao}}]{Pont-2014}%
  \BibitemOpen
  \bibfield  {author} {\bibinfo {author} {\bibfnamefont {S.}~\bibnamefont
  {Courrech~du Pont}}, \bibinfo {author} {\bibfnamefont {C.}~\bibnamefont
  {Narteau}}, \ and\ \bibinfo {author} {\bibfnamefont {X.}~\bibnamefont
  {Gao}},\ }\href@noop {} {\bibfield  {journal} {\bibinfo  {journal} {Geology}\
  }\textbf {\bibinfo {volume} {42}},\ \bibinfo {pages} {743} (\bibinfo {year}
  {2014})}\BibitemShut {NoStop}%
\bibitem [{\citenamefont {Rubin}\ and\ \citenamefont
  {Hunter}(1987)}]{Rubin-1987}%
  \BibitemOpen
  \bibfield  {author} {\bibinfo {author} {\bibfnamefont {D.~M.}\ \bibnamefont
  {Rubin}}\ and\ \bibinfo {author} {\bibfnamefont {R.~E.}\ \bibnamefont
  {Hunter}},\ }\href@noop {} {\bibfield  {journal} {\bibinfo  {journal}
  {Science}\ }\textbf {\bibinfo {volume} {237}},\ \bibinfo {pages} {276}
  (\bibinfo {year} {1987})}\BibitemShut {NoStop}%
\bibitem [{\citenamefont {Rubin}\ and\ \citenamefont
  {Ikeda}(1990)}]{Rubin-1990}%
  \BibitemOpen
  \bibfield  {author} {\bibinfo {author} {\bibfnamefont {D.~M.}\ \bibnamefont
  {Rubin}}\ and\ \bibinfo {author} {\bibfnamefont {H.}~\bibnamefont {Ikeda}},\
  }\href {https://doi.org/10.1111/j.1365-3091.1990.tb00628.x} {\bibfield
  {journal} {\bibinfo  {journal} {Sedimentology}\ }\textbf {\bibinfo {volume}
  {37}},\ \bibinfo {pages} {673} (\bibinfo {year} {1990})}\BibitemShut
  {NoStop}%
\bibitem [{\citenamefont {Taniguchi}\ \emph {et~al.}(2012)\citenamefont
  {Taniguchi}, \citenamefont {Endo},\ and\ \citenamefont
  {Sekiguchi}}]{Taniguchi-2012}%
  \BibitemOpen
  \bibfield  {author} {\bibinfo {author} {\bibfnamefont {K.}~\bibnamefont
  {Taniguchi}}, \bibinfo {author} {\bibfnamefont {N.}~\bibnamefont {Endo}}, \
  and\ \bibinfo {author} {\bibfnamefont {H.}~\bibnamefont {Sekiguchi}},\ }\href
  {https://doi.org/10.1016/j.geomorph.2012.08.019} {\bibfield  {journal}
  {\bibinfo  {journal} {Geomorphology}\ }\textbf {\bibinfo {volume} {179}},\
  \bibinfo {pages} {286} (\bibinfo {year} {2012})}\BibitemShut {NoStop}%
\bibitem [{\citenamefont {Reffet}\ \emph {et~al.}(2010)\citenamefont {Reffet},
  \citenamefont {Courrech~du Pont}, \citenamefont {Hersen},\ and\ \citenamefont
  {Douady}}]{Reffet-2010}%
  \BibitemOpen
  \bibfield  {author} {\bibinfo {author} {\bibfnamefont {E.}~\bibnamefont
  {Reffet}}, \bibinfo {author} {\bibfnamefont {S.}~\bibnamefont {Courrech~du
  Pont}}, \bibinfo {author} {\bibfnamefont {P.}~\bibnamefont {Hersen}}, \ and\
  \bibinfo {author} {\bibfnamefont {S.}~\bibnamefont {Douady}},\ }\href
  {https://doi.org/10.1130/G30894.1} {\bibfield  {journal} {\bibinfo  {journal}
  {Geology}\ }\textbf {\bibinfo {volume} {38}},\ \bibinfo {pages} {491}
  (\bibinfo {year} {2010})}\BibitemShut {NoStop}%
\bibitem [{\citenamefont {Parteli}\ \emph {et~al.}(2009)\citenamefont
  {Parteli}, \citenamefont {Dur\'an}, \citenamefont {Tsoar}, \citenamefont
  {Schw\"ammle},\ and\ \citenamefont {Herrmann}}]{Parteli-2009}%
  \BibitemOpen
  \bibfield  {author} {\bibinfo {author} {\bibfnamefont {E.}~\bibnamefont
  {Parteli}}, \bibinfo {author} {\bibfnamefont {O.}~\bibnamefont {Dur\'an}},
  \bibinfo {author} {\bibfnamefont {H.}~\bibnamefont {Tsoar}}, \bibinfo
  {author} {\bibfnamefont {V.}~\bibnamefont {Schw\"ammle}}, \ and\ \bibinfo
  {author} {\bibfnamefont {H.}~\bibnamefont {Herrmann}},\ }\href
  {https://doi.org/10.1073/pnas.0808646106} {\bibfield  {journal} {\bibinfo
  {journal} {PNAS}\ }\textbf {\bibinfo {volume} {106}},\ \bibinfo {pages}
  {22085} (\bibinfo {year} {2009})}\BibitemShut {NoStop}%
\bibitem [{\citenamefont {Parteli}\ and\ \citenamefont
  {Herrmann}(2007)}]{Parteli-2007}%
  \BibitemOpen
  \bibfield  {author} {\bibinfo {author} {\bibfnamefont {E.~J.~R.}\
  \bibnamefont {Parteli}}\ and\ \bibinfo {author} {\bibfnamefont {H.~J.}\
  \bibnamefont {Herrmann}},\ }\href {\doibase 10.1103/PhysRevLett.98.198001}
  {\bibfield  {journal} {\bibinfo  {journal} {Phys. Rev. Lett.}\ }\textbf
  {\bibinfo {volume} {98}},\ \bibinfo {pages} {198001} (\bibinfo {year}
  {2007})}\BibitemShut {NoStop}%
\bibitem [{\citenamefont {Werner}\ and\ \citenamefont
  {Kocurek}(1997)}]{Werner-1997}%
  \BibitemOpen
  \bibfield  {author} {\bibinfo {author} {\bibfnamefont {B.}~\bibnamefont
  {Werner}}\ and\ \bibinfo {author} {\bibfnamefont {G.}~\bibnamefont
  {Kocurek}},\ }\href
  {https://doi.org/10.1130/0091-7613(1997)025<0771:BFDDTT>2.3.CO;2} {\bibfield
  {journal} {\bibinfo  {journal} {Geology}\ }\textbf {\bibinfo {volume} {25}},\
  \bibinfo {pages} {771} (\bibinfo {year} {1997})}\BibitemShut {NoStop}%
\bibitem [{\citenamefont {Gadal}\ \emph {et~al.}(2019)\citenamefont {Gadal},
  \citenamefont {Narteau}, \citenamefont {Courrech~du Pont}, \citenamefont
  {Rozier},\ and\ \citenamefont {Claudin}}]{Gadal-2019}%
  \BibitemOpen
  \bibfield  {author} {\bibinfo {author} {\bibfnamefont {C.}~\bibnamefont
  {Gadal}}, \bibinfo {author} {\bibfnamefont {C.}~\bibnamefont {Narteau}},
  \bibinfo {author} {\bibfnamefont {S.}~\bibnamefont {Courrech~du Pont}},
  \bibinfo {author} {\bibfnamefont {O.}~\bibnamefont {Rozier}}, \ and\ \bibinfo
  {author} {\bibfnamefont {P.}~\bibnamefont {Claudin}},\ }\href {\doibase
  10.1017/jfm.2018.978} {\bibfield  {journal} {\bibinfo  {journal} {Journal of
  Fluid Mechanics}\ }\textbf {\bibinfo {volume} {862}},\ \bibinfo {pages}
  {490–516} (\bibinfo {year} {2019})}\BibitemShut {NoStop}%
\bibitem [{\citenamefont {Werner}(1995)}]{Werner-1995}%
  \BibitemOpen
  \bibfield  {author} {\bibinfo {author} {\bibfnamefont {B.~T.}\ \bibnamefont
  {Werner}},\ }\href
  {https://doi.org/10.1130/0091-7613(1995)023<1107:EDCSAA>2.3.CO;2} {\bibfield
  {journal} {\bibinfo  {journal} {Geology}\ }\textbf {\bibinfo {volume} {23}},\
  \bibinfo {pages} {1107} (\bibinfo {year} {1995})}\BibitemShut {NoStop}%
\bibitem [{\citenamefont {Nishimori}\ and\ \citenamefont
  {Ouchi}(1993)}]{Nishimori-1993}%
  \BibitemOpen
  \bibfield  {author} {\bibinfo {author} {\bibfnamefont {H.}~\bibnamefont
  {Nishimori}}\ and\ \bibinfo {author} {\bibfnamefont {N.}~\bibnamefont
  {Ouchi}},\ }\href {\doibase 10.1103/PhysRevLett.71.197} {\bibfield  {journal}
  {\bibinfo  {journal} {Phys. Rev. Lett.}\ }\textbf {\bibinfo {volume} {71}},\
  \bibinfo {pages} {197} (\bibinfo {year} {1993})}\BibitemShut {NoStop}%
\bibitem [{\citenamefont {Nishimori}\ \emph {et~al.}(1998)\citenamefont
  {Nishimori}, \citenamefont {Yamasaki},\ and\ \citenamefont
  {Andersen}}]{Nishimori-1998}%
  \BibitemOpen
  \bibfield  {author} {\bibinfo {author} {\bibfnamefont {H.}~\bibnamefont
  {Nishimori}}, \bibinfo {author} {\bibfnamefont {M.}~\bibnamefont {Yamasaki}},
  \ and\ \bibinfo {author} {\bibfnamefont {K.~H.}\ \bibnamefont {Andersen}},\
  }\href {https://doi.org/10.1142/S021797929800020X} {\bibfield  {journal}
  {\bibinfo  {journal} {International Journal of Modern Physics B}\ }\textbf
  {\bibinfo {volume} {12}},\ \bibinfo {pages} {257} (\bibinfo {year}
  {1998})}\BibitemShut {NoStop}%
\bibitem [{\citenamefont {Guignier}\ \emph {et~al.}(2013)\citenamefont
  {Guignier}, \citenamefont {Niiya}, \citenamefont {Nishimori}, \citenamefont
  {Lague},\ and\ \citenamefont {Valance}}]{Guignier-2013}%
  \BibitemOpen
  \bibfield  {author} {\bibinfo {author} {\bibfnamefont {L.}~\bibnamefont
  {Guignier}}, \bibinfo {author} {\bibfnamefont {H.}~\bibnamefont {Niiya}},
  \bibinfo {author} {\bibfnamefont {H.}~\bibnamefont {Nishimori}}, \bibinfo
  {author} {\bibfnamefont {D.}~\bibnamefont {Lague}}, \ and\ \bibinfo {author}
  {\bibfnamefont {A.}~\bibnamefont {Valance}},\ }\href@noop {} {\bibfield
  {journal} {\bibinfo  {journal} {Phys. Rev. E}\ }\textbf {\bibinfo {volume}
  {87}},\ \bibinfo {pages} {052206} (\bibinfo {year} {2013})}\BibitemShut
  {NoStop}%
\bibitem [{\citenamefont {Niiya}\ \emph {et~al.}(2010)\citenamefont {Niiya},
  \citenamefont {Awazu},\ and\ \citenamefont {Nishimori}}]{Niiya-2010}%
  \BibitemOpen
  \bibfield  {author} {\bibinfo {author} {\bibfnamefont {H.}~\bibnamefont
  {Niiya}}, \bibinfo {author} {\bibfnamefont {A.}~\bibnamefont {Awazu}}, \ and\
  \bibinfo {author} {\bibfnamefont {H.}~\bibnamefont {Nishimori}},\ }\href
  {https://doi.org/10.1143/JPSJ.79.063002} {\bibfield  {journal} {\bibinfo
  {journal} {Journal of the Physical Society of Japan}\ }\textbf {\bibinfo
  {volume} {79}},\ \bibinfo {pages} {063002} (\bibinfo {year}
  {2010})}\BibitemShut {NoStop}%
\bibitem [{\citenamefont {Niiya}\ \emph {et~al.}(2012)\citenamefont {Niiya},
  \citenamefont {Awazu},\ and\ \citenamefont {Nishimori}}]{Niiya-2012}%
  \BibitemOpen
  \bibfield  {author} {\bibinfo {author} {\bibfnamefont {H.}~\bibnamefont
  {Niiya}}, \bibinfo {author} {\bibfnamefont {A.}~\bibnamefont {Awazu}}, \ and\
  \bibinfo {author} {\bibfnamefont {H.}~\bibnamefont {Nishimori}},\ }\href
  {https://link.aps.org/doi/10.1103/PhysRevLett.108.158001} {\bibfield
  {journal} {\bibinfo  {journal} {Phys. Rev. Lett.}\ }\textbf {\bibinfo
  {volume} {108}},\ \bibinfo {pages} {158001} (\bibinfo {year}
  {2012})}\BibitemShut {NoStop}%
\bibitem [{\citenamefont {Momiji}\ and\ \citenamefont
  {Warren}(2000)}]{Momiji-2000}%
  \BibitemOpen
  \bibfield  {author} {\bibinfo {author} {\bibfnamefont {H.}~\bibnamefont
  {Momiji}}\ and\ \bibinfo {author} {\bibfnamefont {A.}~\bibnamefont
  {Warren}},\ }\href {\doibase
  10.1002/1096-9837(200009)25:10<1069::AID-ESP117>3.0.CO;2-D} {\bibfield
  {journal} {\bibinfo  {journal} {Earth Surface Processes and Landforms}\
  }\textbf {\bibinfo {volume} {25}},\ \bibinfo {pages} {1069} (\bibinfo {year}
  {2000})}\BibitemShut {NoStop}%
\bibitem [{Note1()}]{Note1}%
  \BibitemOpen
  \bibinfo {note} {{ Since $h$, $x$, and $y$ have the same dimension of unit,
  there are actually three possible ways to interpret the numerical values
  depending on the choice of two quantities out of the three to be assigned to
  the same unit. We prefer $h$ and $x$ to be in the same unit because the
  geometrical parameters $A$, $B$, and $C$ do not depend on the constants of
  Eq.(\ref {unit_system}). In this choice, however, the slant angle $\chi $
  depends on $\protect \sqrt {D_u/q_c}$ as in Eq.(\ref {tilde_chi}).
  }}\BibitemShut {NoStop}%
\bibitem [{Note2()}]{Note2}%
  \BibitemOpen
  \bibinfo {note} {{ In the following, numerical values for the slant angle are
  always given by $\protect \mathaccentV {tilde}07E\chi $ defined in Eq.(\ref
  {tilde_chi}), but the slant angle in the mathematical expressions should be
  the original slant angle $\chi $. }}\BibitemShut {NoStop}%
\bibitem [{Note3()}]{Note3}%
  \BibitemOpen
  \bibinfo {note} {It is interesting to see that the average growth rate for
  the bimodal wind can be larger than the growth rate for the unidirectional
  wind. This means that the largest absolute value of the eigenvalue of the
  evolution matrix $\protect \mathaccentV {hat}05E\Lambda _{\protect \rm
  trans}$ is larger than the product of the largest absolute values of the
  eigenvalue of each factor matrix in Eq.(\ref {Lambda_trans}).}\BibitemShut
  {Stop}%
\bibitem [{\citenamefont {Tsoar}\ \emph {et~al.}(2004)\citenamefont {Tsoar},
  \citenamefont {Blumberg},\ and\ \citenamefont {Stoler}}]{Tsoar-2004}%
  \BibitemOpen
  \bibfield  {author} {\bibinfo {author} {\bibfnamefont {H.}~\bibnamefont
  {Tsoar}}, \bibinfo {author} {\bibfnamefont {D.~G.}\ \bibnamefont {Blumberg}},
  \ and\ \bibinfo {author} {\bibfnamefont {Y.}~\bibnamefont {Stoler}},\ }\href
  {https://doi.org/10.1016/S0169-555X(03)00161-2} {\bibfield  {journal}
  {\bibinfo  {journal} {Geomorphology}\ }\textbf {\bibinfo {volume} {57}},\
  \bibinfo {pages} {293} (\bibinfo {year} {2004})}\BibitemShut {NoStop}%
\bibitem [{\citenamefont {Pye}\ and\ \citenamefont {Tsoar}(2009)}]{Pye-2009}%
  \BibitemOpen
  \bibfield  {author} {\bibinfo {author} {\bibfnamefont {K.}~\bibnamefont
  {Pye}}\ and\ \bibinfo {author} {\bibfnamefont {H.}~\bibnamefont {Tsoar}},\
  }in\ \href@noop {} {\emph {\bibinfo {booktitle} {Aeolian Sand and Sand
  Dunes}}}\ (\bibinfo  {publisher} {Springer},\ \bibinfo {address} {Berlin,
  Heidelberg},\ \bibinfo {year} {2009})\ Chap.~\bibinfo {chapter} {6}, pp.\
  \bibinfo {pages} {175--253}\BibitemShut {NoStop}%
\bibitem [{\citenamefont {Momiji}(2001)}]{Momiji-2001}%
  \BibitemOpen
  \bibfield  {author} {\bibinfo {author} {\bibfnamefont {H.}~\bibnamefont
  {Momiji}},\ }\emph {\bibinfo {title} {Mathematical modelling of the dynamics
  and morphology of aeolian dunes and dune fields}},\ \href
  {https://inis.iaea.org/search/search.aspx?orig_q=RN:34022958} {Ph.D.
  thesis},\ \bibinfo  {school} {University College London} (\bibinfo {year}
  {2001})\BibitemShut {NoStop}%
\bibitem [{Note4()}]{Note4}%
  \BibitemOpen
  \bibinfo {note} {{ From the numerical estimates of the trapping efficiency
  $T_e$ (Fig.3.3 at p.50 of \cite {Warren}) and the friction velocity $v^*$
  (Fig.4.5 at p.88 of \cite {Momiji-2001}), we obtain the estimates $h_c\sim 5$
  and $u^*\sim $ 0.5 m/s. For this value of the friction velocity, the mass
  flux at the crest $q_{\protect \rm mass}$ can be estimated as $q_{\protect
  \rm mass}\sim 0.01$ kg/m s from Fig.1.8 in p.28 of \cite {Warren} . The
  volume flux $q_c$ is given by $q_c=q_{\protect \rm mass}/\rho $ with the mass
  density of sand $\rho \approx 1.7\times 10^{3} {\protect \rm \protect
  \tmspace +\thinmuskip {.1667em} kg\protect \tmspace +\thinmuskip {.1667em}
  m^{-3}}$.}}\BibitemShut {Stop}%
\bibitem [{\citenamefont {Schw\"ammle}\ and\ \citenamefont
  {Herrmann}(2005)}]{Schwammle-2005}%
  \BibitemOpen
  \bibfield  {author} {\bibinfo {author} {\bibfnamefont {V.}~\bibnamefont
  {Schw\"ammle}}\ and\ \bibinfo {author} {\bibfnamefont {H.}~\bibnamefont
  {Herrmann}},\ }\href {https://doi.org/10.1140/epje/e2005-00007-0} {\bibfield
  {journal} {\bibinfo  {journal} {Eur. Phys. J. E}\ }\textbf {\bibinfo {volume}
  {16}},\ \bibinfo {pages} {57} (\bibinfo {year} {2005})}\BibitemShut {NoStop}%
\bibitem [{\citenamefont {Lorenz}\ \emph {et~al.}(2006)\citenamefont {Lorenz},
  \citenamefont {Wall}, \citenamefont {Radebaugh}, \citenamefont {Boubin},
  \citenamefont {Reffet}, \citenamefont {Janssen}, \citenamefont {Stofan},
  \citenamefont {Lopes}, \citenamefont {Kirk}, \citenamefont {Elachi},
  \citenamefont {Lunine}, \citenamefont {Mitchell}, \citenamefont {Paganelli},
  \citenamefont {Soderblom}, \citenamefont {Wood}, \citenamefont {Wye},
  \citenamefont {Zebker}, \citenamefont {Anderson}, \citenamefont {Ostro},
  \citenamefont {Allison}, \citenamefont {Boehmer}, \citenamefont {Callahan},
  \citenamefont {Encrenaz}, \citenamefont {Ori}, \citenamefont {Francescetti},
  \citenamefont {Gim}, \citenamefont {Hamilton}, \citenamefont {Hensley},
  \citenamefont {Johnson}, \citenamefont {Kelleher}, \citenamefont {Muhleman},
  \citenamefont {Picardi}, \citenamefont {Posa}, \citenamefont {Roth},
  \citenamefont {Seu}, \citenamefont {Shaffer}, \citenamefont {Stiles},
  \citenamefont {Vetrella}, \citenamefont {Flamini},\ and\ \citenamefont
  {West}}]{Lorenz-2006}%
  \BibitemOpen
  \bibfield  {author} {\bibinfo {author} {\bibfnamefont {R.~D.}\ \bibnamefont
  {Lorenz}}, \bibinfo {author} {\bibfnamefont {S.}~\bibnamefont {Wall}},
  \bibinfo {author} {\bibfnamefont {J.}~\bibnamefont {Radebaugh}}, \bibinfo
  {author} {\bibfnamefont {G.}~\bibnamefont {Boubin}}, \bibinfo {author}
  {\bibfnamefont {E.}~\bibnamefont {Reffet}}, \bibinfo {author} {\bibfnamefont
  {M.}~\bibnamefont {Janssen}}, \bibinfo {author} {\bibfnamefont
  {E.}~\bibnamefont {Stofan}}, \bibinfo {author} {\bibfnamefont
  {R.}~\bibnamefont {Lopes}}, \bibinfo {author} {\bibfnamefont
  {R.}~\bibnamefont {Kirk}}, \bibinfo {author} {\bibfnamefont {C.}~\bibnamefont
  {Elachi}}, \bibinfo {author} {\bibfnamefont {J.}~\bibnamefont {Lunine}},
  \bibinfo {author} {\bibfnamefont {K.}~\bibnamefont {Mitchell}}, \bibinfo
  {author} {\bibfnamefont {F.}~\bibnamefont {Paganelli}}, \bibinfo {author}
  {\bibfnamefont {L.}~\bibnamefont {Soderblom}}, \bibinfo {author}
  {\bibfnamefont {C.}~\bibnamefont {Wood}}, \bibinfo {author} {\bibfnamefont
  {L.}~\bibnamefont {Wye}}, \bibinfo {author} {\bibfnamefont {H.}~\bibnamefont
  {Zebker}}, \bibinfo {author} {\bibfnamefont {Y.}~\bibnamefont {Anderson}},
  \bibinfo {author} {\bibfnamefont {S.}~\bibnamefont {Ostro}}, \bibinfo
  {author} {\bibfnamefont {M.}~\bibnamefont {Allison}}, \bibinfo {author}
  {\bibfnamefont {R.}~\bibnamefont {Boehmer}}, \bibinfo {author} {\bibfnamefont
  {P.}~\bibnamefont {Callahan}}, \bibinfo {author} {\bibfnamefont
  {P.}~\bibnamefont {Encrenaz}}, \bibinfo {author} {\bibfnamefont {G.~G.}\
  \bibnamefont {Ori}}, \bibinfo {author} {\bibfnamefont {G.}~\bibnamefont
  {Francescetti}}, \bibinfo {author} {\bibfnamefont {Y.}~\bibnamefont {Gim}},
  \bibinfo {author} {\bibfnamefont {G.}~\bibnamefont {Hamilton}}, \bibinfo
  {author} {\bibfnamefont {S.}~\bibnamefont {Hensley}}, \bibinfo {author}
  {\bibfnamefont {W.}~\bibnamefont {Johnson}}, \bibinfo {author} {\bibfnamefont
  {K.}~\bibnamefont {Kelleher}}, \bibinfo {author} {\bibfnamefont
  {D.}~\bibnamefont {Muhleman}}, \bibinfo {author} {\bibfnamefont
  {G.}~\bibnamefont {Picardi}}, \bibinfo {author} {\bibfnamefont
  {F.}~\bibnamefont {Posa}}, \bibinfo {author} {\bibfnamefont {L.}~\bibnamefont
  {Roth}}, \bibinfo {author} {\bibfnamefont {R.}~\bibnamefont {Seu}}, \bibinfo
  {author} {\bibfnamefont {S.}~\bibnamefont {Shaffer}}, \bibinfo {author}
  {\bibfnamefont {B.}~\bibnamefont {Stiles}}, \bibinfo {author} {\bibfnamefont
  {S.}~\bibnamefont {Vetrella}}, \bibinfo {author} {\bibfnamefont
  {E.}~\bibnamefont {Flamini}}, \ and\ \bibinfo {author} {\bibfnamefont
  {R.}~\bibnamefont {West}},\ }\href {\doibase 10.1126/science.1123257}
  {\bibfield  {journal} {\bibinfo  {journal} {Science}\ }\textbf {\bibinfo
  {volume} {312}},\ \bibinfo {pages} {724} (\bibinfo {year}
  {2006})}\BibitemShut {NoStop}%
\bibitem [{\citenamefont {Tokano}(2008)}]{tokano-2008}%
  \BibitemOpen
  \bibfield  {author} {\bibinfo {author} {\bibfnamefont {T.}~\bibnamefont
  {Tokano}},\ }\href {\doibase https://doi.org/10.1016/j.icarus.2007.10.007}
  {\bibfield  {journal} {\bibinfo  {journal} {Icarus}\ }\textbf {\bibinfo
  {volume} {194}},\ \bibinfo {pages} {243 } (\bibinfo {year}
  {2008})}\BibitemShut {NoStop}%
\end{thebibliography}%
\end{document}